\documentclass[acmsmall]{acmart}

\AtBeginDocument{%
  \providecommand\BibTeX{{%
    \normalfont B\kern-0.5em{\scshape i\kern-0.25em b}\kern-0.8em\TeX}}}

\setcopyright{acmcopyright}
\copyrightyear{2023}
\acmYear{2023}
\acmDOI{XXXXXXX.XXXXXXX}

\acmJournal{JACM}
\acmVolume{000}
\acmNumber{000}
\acmArticle{000}
\acmMonth{000}

\usepackage{graphicx}%
\usepackage{multirow}%
\usepackage{amsthm}%
\usepackage{mathrsfs}%
\usepackage[title]{appendix}%
\usepackage{xcolor}%
\usepackage{textcomp}%
\usepackage{manyfoot}%
\usepackage{algorithm}%
\usepackage{algorithmicx}%
\usepackage{algpseudocode}%
\usepackage{listings}%
\usepackage{subfigure}
\usepackage{bm}
\usepackage{enumitem}

\begin{document}

\title{BHEISR: Nudging from Bias to Balance - Promoting Belief Harmony by Eliminating Ideological Segregation in Knowledge-based Recommendations}

\author{Mengyan Wang}
\email{mengyan.wang@autuni.ac.nz}
\affiliation{%
  \institution{Auckland University of Technology}
  \city{Auckland}
  \country{New Zealand}
}

\author{Yuxuan Hu}
\email{yuxuan.hu@utas.edu.au}
\affiliation{%
  \institution{University of Tasmania}
  \city{Hobart}
  \country{Australia}
}

\author{Zihan Yuan}
\email{zyuan0@utas.edu.au}
\affiliation{%
  \institution{University of Tasmania}
  \city{Hobart}
  \country{Australia}
}

\author{Chenting Jiang}
\email{chenting.jiang@utas.edu.au}
\affiliation{%
  \institution{University of Tasmania}
  \city{Hobart}
  \country{Australia}
}

\author{Weihua Li}
\authornote{Corresponding author}
\email{weihua.li@aut.ac.nz}
\affiliation{%
  \institution{Auckland University of Technology}
  \city{Auckland}
  \country{New Zealand}
}

\author{Shiqing Wu}
\email{Shiqing.Wu@uts.edu.au}
\affiliation{%
  \institution{University of Technology Sydney}
  \city{Sydney}
  \country{Australia}
}

\author{Quan Bai}
\email{quan.bai@utas.edu.au}
\affiliation{%
  \institution{University of Tasmania}
  \city{Hobart}
  \country{Australia}
}

\renewcommand{\shortauthors}{Wang et al.}

\begin{abstract}

In the realm of personalized recommendation systems, the increasing concern is the amplification of belief imbalance and user biases, a phenomenon primarily attributed to the filter bubble. Addressing this critical issue, we introduce an innovative intermediate agency (BHEISR) between users and existing recommendation systems to attenuate the negative repercussions of the filter bubble effect in extant recommendation systems. The main objective is to strike a belief balance for users while minimizing the detrimental influence caused by filter bubbles. The BHEISR model amalgamates principles from nudge theory while upholding democratic and transparent principles. It harnesses user-specific category information to stimulate curiosity, even in areas users might initially deem uninteresting. By progressively stimulating interest in novel categories, the model encourages users to broaden their belief horizons and explore the information they typically overlook. Our model is time-sensitive and operates on a user feedback loop. It utilizes the existing recommendation algorithm of the model and incorporates user feedback from the prior time frame. This approach endeavors to transcend the constraints of the filter bubble, enrich recommendation diversity, and strike a belief balance among users while also catering to user preferences and system-specific business requirements. To validate the effectiveness and reliability of the BHEISR model, we conducted a series of comprehensive experiments with real-world datasets. These experiments compared the performance of the BHEISR model against several baseline models using nearly 200 filter bubble-impacted users as test subjects. Our experimental results conclusively illustrate the superior performance of the BHEISR model in mitigating filter bubbles and balancing user perspectives.

\end{abstract}




\maketitle

\section{Introduction}

The presence of filter bubbles in online recommendations has raised concerns about the potential negative impact on society and individuals \cite{bawden2020information}. Filter bubbles refer to users being fragmented by continuously and passively receiving homogeneous items due to over-specific algorithm recommendations and personal preferences \cite{pardos2019combating}. In the era of big data, recommendation systems play a crucial role in shaping users' perspectives and access to diverse information \cite{naumov2019deep}. However, the personalization recommendation algorithms are reckoned as the filter bubbles' main culprit. Under the filter bubbles, each person has a sole information universe and is alone in their specific world \cite{michiels2022filter}. These filter bubble-existed recommendation systems often decrease the diversity of users' beliefs, leading to bias and ideological segregation, reinforcing users' existing beliefs and limiting their exposure to alternative viewpoints. This phenomenon not only hinders the users' formation of a well-rounded understanding but also exacerbates societal polarization \cite{ekstrom2021self}.

To address the prevailing issue of societal polarization in online social networks, existing research primarily deploys two types of strategies embedded in recommendation systems: algorithm-focused and human-focused \cite{alatawi2021survey}. The algorithm-focused strategies combat filter bubbles by advocating for content diversity during the in-processing and post-processing stages \cite{alatawi2021survey}. Methods such as explanation-based diversity recommendation \cite{yu2009takes}, community-aware model \cite{grossetti2019community}, category-based diversification algorithm \cite{lunardi2020metric}, the Diversified GNN-based Recommender system (DGRec) \cite{yang2023dgrec}, and the graph-based user-item interaction method \cite{li2023breaking} are employed. These strategies, particularly those based on graph-based recommendations, offer invaluable insights into user preferences and category diversity. However, their heavy reliance on algorithmic engines to curate recommended content inadvertently neglects the integral role of human decision-making processes.
In contrast, human-focused strategies privilege individual agency over algorithmic engines when addressing the problem of filter bubbles \cite{alatawi2021survey}. For instance, nudging-based recommendations \cite{jesse2021digital,joachim2022nudge,tangruamsubcaregraph}. Unlike algorithm-focused models, nudge recommendations replace directly serving up targeted information to users, but adopt an indirect method of influencing individuals' decisions and behaviors \cite{beshears2020nudging,blumenthal2018nudge}. Despite this, existing nudging-based models often grapple with effectively and explainable expanding users' interest into categories previously shown no interest. Therefore, this study amalgamates the benefits of both graph-based algorithmic strategies and nudging-based recommendation systems to provide a responsible system to effectively mitigate the impact of the filter bubble.

\begin{figure}[htbp]
		\centering
		\includegraphics[width=0.35\textwidth]{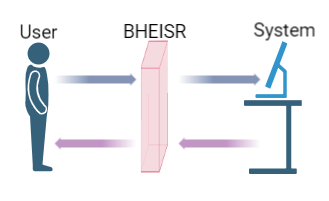}
    \caption{BHEISR: An intermediate agency between users and recommendation systems.}
	 \label{fig:user_turn}
\end{figure}

This research aims to address the challenge posed by the filter bubble phenomenon and promote belief harmony among users by implementing nudge recommendations. We propose a novel recommendation approach, BHEISR, which functions as an intermediary between existing recommendation systems and user beliefs, as depicted in Figure \ref{fig:user_turn}. BHEISR is designed to mitigate the adverse effects of preference-based recommendation algorithms in systems susceptible to filter bubbles. With regard to users who are already impacted by filter bubbles, the model employs nudging techniques to encourage them to step outside their personal awareness comfort zones, fostering a more balanced representation of users' beliefs.

In pursuit of our objective, we initially propose a Filter Bubbles Detection Model based on Multi-faceted Reasoning (FBDMR). The model utilizes a bipartite approach to ascertain users who are influenced by the filter bubble and to identify the systems implementing user preference-based recommendation algorithms. The filter bubble impacts users, resulting in unbalanced beliefs, signifying a user's narrow or disinterested perspectives on a specific type of information (considered as a category) when they show extreme interest in another. Subsequently, we develop a recommendation prompt path incorporating these two categories (extremely interested or normal) by investigating more interconnected nodes between them. The BHEISR model employs nudge recommendations, embodying libertarian paternalism and adhering to principles of transparency. The aim is to subtly guide users towards exploring diverse viewpoints and questioning their preconceived notions via this generated prompt path. Throughout this process, the Generate Artificial Intelligence (GAI) model plays an essential role in further extracting abundant contextual information in the realm of big data.

The key contributions of this research can be categorized into four aspects.

\begin{itemize}

    \item Firstly, we propose a novel model, BHEISR, to alleviate the adverse impacts of the filter bubble. The model balances users' interests and the recommendation system. Also, BHEISR equilibrates existing recommendation lists while offering democratic and interpretable suggestions to users. Furthermore, it promptly gathers user feedback, attaining belief harmony for users. 

    \item Secondly, our research introduces a Filter Bubbles Detection Model based on Multi-faceted Reasoning (FBDMR). This model detects and identifies users affected by filter bubbles and recommendation systems where filter bubbles exist. With diverse methodologies, we delve into a fine-grained analysis of the existence and effects of filter bubbles from various detection aspects.
    
    \item Thirdly, we harness the power of nudging techniques to gently guide users towards broadening their interests and nurturing belief harmony. The process of nudge recommendation adheres to the principles of libertarian paternalism, transparency, and democracy, which enhances users' comprehension of the recommendations and boosts their confidence. 
    
    \item Lastly, we employ the emerging GAI techniques within BHEISR to produce prompt path-based textual content for users and to solicit their feedback for updating belief graphs. The GAI model expands these belief graphs, deeply investigating the relationship between users' preferences and non-preferences, and conveys a broader range of semantic information to users, thereby enhancing diversity.
    
\end{itemize}

The remainder of this paper is organized as follows. In Section 2, we provide a comprehensive review of related work on recommendation systems, user selective exposure, filter bubbles, and nudging techniques. Section 3 describes our formal definitions, user belief networks, and category correlation graph construction. Section 4 presents the methodology and framework employed in our research. In Section 5, we discuss the experimental setup. Results and analysis are also presented in this section, followed by a discussion of the findings in Section 6. Finally, Section 7 concludes the paper and outlines potential avenues for future research.

\section{Related Works}

Despite their popularity and ability to provide users with tailored content suggestions, personalized recommendation systems inadvertently create a significant challenge: filter bubbles \cite{patro2020fairrec}. Filter bubbles constrain users' exposure to various perspectives and information, thus potentially leading to belief biases and societal fragmentation \cite{lunardi2020metric}. To mitigate this concern, an increasing number of researchers and practitioners have turned their attention towards dismantling filter bubbles, fostering diversity and democracy in recommender systems, and facilitating users' belief harmony.

In this section, we comprehensively review the relevant research works, deliberate on the filter bubble issue, explore the diversification of recommender systems, and review existing works on nudge recommendations. Additionally, we will highlight the contributions of this study. 

\subsection{Filter Bubbles and Ideological Segregation}

\subsubsection{Preference-based Recommendation Systems}

Conventional recommendation systems prioritize the generalization of user preference, implying that these systems often recommend items to users based on their specific preferences and behaviors \cite{guo2020attentional}. Techniques such as Collaborative Filtering (CF) \cite{wang2021robust}, Content-Based filtering (CB) \cite{reddy2019content}, rule-based methods \cite{wu2020scalability}, or hybrid models \cite{afoudi2021hybrid} are commonly employed to analyze users' preferences and past behaviors. The recommendation system then suggests content that aligns closely with user preferences to enhance user satisfaction and engagement. However, this approach based on user preference may exacerbate the filter bubble issue, leading to ideological isolation and user bias. For example, Bryant et al. demonstrate that the YouTube algorithm, representative of a preference recommendation algorithm, exhibits a marked bias towards right-leaning political videos, including those espousing racist views propagated by the alt-right community \cite{bryant2020youtube}. It has become important to address the limitations of current preference recommendations, boost the diversity of suggestions, and actualize users' belief harmony.

\subsubsection{Mitigating Filter Bubble Effects}

In contrast to the echo chamber, filter bubbles emphasize on the constraints of preference recommendation algorithms \cite{kitchens2020understanding}. Dahlgren et al. introduce the term "internet filters" to encapsulate the phenomenon of filter bubbles, which can entail numerous repercussions for users, including a narrowed focus on personal interests, substantial reinforcement of confirmation bias, reduced curiosity, decreased exposure to diverse ideas and people, compromised understanding of the world, and a skewed perception of reality \cite{dahlgren2021critical}.

Addressing the negative effects of filter bubbles necessitates recognizing various challenges, with the algorithmic bias being particularly noteworthy. Chen et al. highlight that the emergence of recommendation algorithm bias amplifies the experimental nature of user behavior data as opposed to observational \cite{chen2023bias}. %This preference-based experimental data yields more preference-oriented recommendation lists, further solidifies users' preferences, and fosters a self-contained ecosystem for each user \cite{chen2023bias}. 
Additionally, Dahlgren et al. examine the recommendation algorithm bias and broaden the concept of bias into two facets \cite{dahlgren2021critical}. One form of bias originates from the recommendation algorithm, while the other stems from users' behaviors. %In our research, we employ a similar concept to detect filter bubbles in two manners, through the identification of unbalanced users' beliefs and the detection of recommendation algorithm bias.
Aside from algorithmic bias, another challenge in mitigating filter bubbles lies in their elusive nature \cite{lunardi2020metric}. Users are often unaware of the homogenized world due to the imperceptible filter bubble effect. Namely, they remain unaware that their viewpoint differs from others in the same situation. %To counteract this issue, we will employ explainable methodologies, visible users' belief graphs, and nudge techniques, thereby rendering our model's entire process understandable for users.

The growing influence of filter bubbles has raised increased concerns among researchers. A well-crafted recommendation system usually offers high accuracy while also promoting diversity; systems oriented solely towards accuracy may inevitably lead to filter bubble effects \cite{yang2023dgrec}. Contemporary research works propose several strategies for breaking filter bubbles by enhancing the diversity of recommendations. The research works of addressing filter bubbles with and without knowledge graphs, as well as selective exposure detection, are reviewed as follows.

\textbf{Research beyond the scope of the knowledge graph.} Early diversification algorithms, initially designed for recommendation systems, sought to augment category diversification \cite{castells2021novelty}. Su et al. propose the category-inspired diversity recommendation algorithm to integrate the dissimilarity of a user's item features, considering the taxonomic tree to widen and enhance the user's interests \cite{su2020diversifying}. %This approach involves the attributes of the suggested items to extend and generalize user preferences \cite{su2020diversifying}.
Yu et al. introduce an attribute-inspired, explanation-based diversity recommendation system, which achieves diversity in recommendation lists from two angles, i.e., item-based and collaborative filtering strategy-based recommendations \cite{yu2009takes}. %Users can receive recommendation items based on their preferences and highly rated items from their group members \cite{yu2009takes}. This methodology effectively fulfills the diversity objective to a certain degree.
Similarly, Grossetti et al. extend the scope from groups to communities, where a community-aware model is proposed to identify Twitter communities and compute their category-based similarities \cite{grossetti2021reducing}. %By integrating a re-ranking model into the recommendation algorithm, the recommendation diversity is enhanced. This model also employs community similarity as a foundation, merging community evaluation scores with recommendation item scores to augment recommendation diversity \cite{grossetti2021reducing}.
Moreover, Lunardi et al. propose a category-based diversification algorithm to diminish the filter bubble effect, ensuring that recommended items align with the user's interests while maintaining a degree of diversity in content and features\cite{lunardi2020metric}. %This deep neural network-based recommender deploys a Maximal Marginal Relevance (MMR) algorithm to ensure that recommended items align with the user's interests while maintaining a degree of diversity in content and features. This strategy prevents the recommendation of excessively similar items and widens the user's choices and opportunities for discovering new content. They introduce the Homogeneity Level (HL) to assess the effectiveness of their proposed system. 

\textbf{Research in the scope of the knowledge graph.} While the aforementioned research models have significantly contributed to alleviating filter bubbles and enhancing recommendation diversity, many researchers also advocate for the critical role of knowledge graph-based recommendation algorithms. These algorithms not only mitigate data sparsity and cold start issues, but they also add an essential interpretability factor to recommendation systems \cite{guo2020survey}.
Yang et al. introduced the Diversified GNN-based Recommender System (DGRec), a graph-based recommendation system built on GNN, augmenting the diversity of recommended lists by improving the embedding generation process \cite{yang2023dgrec}. %The DGRec intends to augment the diversity of recommended lists by improving the embedding generation process. It incorporates three components, i.e., submodular neighbor selection, layer attention, and loss reweighting. By incorporating these components into GNN, DGRec maintains high diversity and comparable accuracy with state-of-the-art GNN-based recommender systems \cite{yang2023dgrec}.
Additionally, Li et al. adopt a graph-based methodology by constructing a user-item interaction graph for data analysis to examine the existence of a centralized recommendation phenomenon \cite{li2023breaking}. %A generic simulation framework is utilized to extrapolate the recommender system's procedure, forming a "user in control" loop. To mitigate the issue of filter bubbles under the feedback loop, they propose a reinforcement learning-based method that selects connections between different communities for the recommendation list \cite{li2023breaking}.
In contrast to these methods, our model transcends the traditional approach of generating marginally varied items based on user preferences for diversity. We prioritize incrementally stimulating users' interest in items they may initially disregard without altering existing recommender system algorithms. The proposed novel approach aims to counteract the filter bubble effect by considering user interest and disinterest beliefs, i.e., an aspect that has received very limited attention from researchers.

\textbf{Detection of Selective Exposure.} Belief bias in reasoning refers to individuals' tendency to favor conclusions that align with their pre-existing beliefs \cite{calvillo2020ideological}. This phenomenon is intimately connected with the formation of online filter bubbles, in which users tend to accept information that confirms their viewpoints and interests while rejecting alternative perspectives that challenge their beliefs \cite{resnick2013bursting, citation-0}. %Within these filter bubbles, users tend to accept information that confirms their viewpoints and interests, while rejecting alternative perspectives that challenge their beliefs. In our research, we define belief bias as selective exposure, which refers to the deliberate act of seeking information that aligns with one's preconceived notions while avoiding dissenting viewpoints. 
Existing methods proposed for selective exposure detection include Information Source Diversity Analysis (ISDA) \cite{lunardi2020metric}, User Interaction Pattern Analysis (UIPA) \cite{wang2022user}, Reinforcement Learning Methods (RLM) \cite{li2023breaking}, and Social Network Analysis (SNA) \cite{wang2022user,chitra2020analyzing}. Considering the limited interpretability of RLM and the focus of SNA on alleviating echo-chamber effects rather than filter bubbles, our research concentrates on investigating selective exposure detection algorithms based on ISDA and UIPA. ISDA includes various detection metrics such as topology metrics and homophily metrics \cite{lunardi2020metric}. Likewise, UIPA includes several established detection metrics, including the coverage algorithm and the Majority Category Domination (MCD) algorithm \cite{wang2022user}.
Drawing inspiration from these metrics, we propose the FBDMR model for dual verification of the authenticity of the filter bubble phenomenon, having the concept of 'Entropy' \cite{xiao2019multi,zhou2021generalized,wang2019new} included to substantiate the existence of filter bubbles.

\subsection{Nudge Techniques and Recommendations}

A nudge is a non-coercive intervention designed to influence behavior by modifying the context in which choices are made \cite{blumenthal2018nudge,beshears2020nudging}. %It avoids resorting to mandatory rules or constraints. 
This form of such an intervention is usually transparent and optional, enabling individuals to better understand their choice consequences and to boost the likelihood of beneficial decision-making \cite{beshears2020nudging}. The core idea behind a nudge is to exploit individuals' beliefs and behavioral biases through various design strategies, directing them towards more favorable outcomes without constraining their freedom of choice \cite{blumenthal2018nudge}.
Given the widespread acceptance of nudging, an increasing number of researchers in recommendation systems are beginning to investigate the impact and effectiveness of nudges in diverse domains of recommendation systems. However, the majority introduces nudging recommendations from an AI-deprived perspective, implying a substantial absence or lack of AI technology in their research context. For example, Jesse et al. consolidate 87 nudging mechanisms at this AI-deprived level, including alterations in font size, the reputation of the messenger, and the visibility of information \cite{jesse2021digital}.
Joachim et al. propose a platform empowered by AI designed to nudge, influence, and guide the behavior of individuals with diabetes \cite{joachim2022nudge}. To enhance the transparency of the nudging process and offer greater interpretability to the interventions, Tangruamsub et al. incorporate a knowledge graph into the nudge-based recommendation system to analyze users' preferences more thoroughly and make more precise recommendations \cite{tangruamsubcaregraph}. %They analyze users' nudging interest graphs, for instance, preferences for louder message alerts, to comprehend their preferences and employ knowledge graphs to uncover potential nudging interests. The goal is to inspire users to adopt a healthier lifestyle through behavioral nudging. 
Furthermore, Sitar et al. propose an automated recommendation system that integrates managers' priorities and user feedback, utilizing knowledge graph structures, to organize items based on descending order of priority, known as nudge concepts \cite{sitar2021knowledge}.

The recommendation systems mentioned previously have revealed the importance of establishing a knowledge graph-based nudging recommendation system. %They have shown the interpretability and scalability exhibited by knowledge graphs in the recommendation process, as well as the transparency and non-coercive nature of nudging within recommendation systems. 
However, these models are developed on the principle of privileging user preferences without recognizing the significance of transforming user disinterest perceptions into acceptable items. Different from the existing approaches, in this research, we strive to amplify recommendation diversity to augment user belief and harmony. The objective is to present items that users might initially dislike or reject in a non-coercive manner, guiding user perceptions from one end of the graph (items congruent with user preferences) to the other end (items less favored by users), enabling a transition in user belief from bias to balance.

%\subsection{Summary}

%In this section, we discussed filter bubbles in recommendation systems, challenges in addressing them, current research on diversifying recommendations, and the use of nudge techniques. 

%Different from the existing approaches, our innovative research combines equitable diversity recommendations and filter bubble detection for belief harmony. It treats objects accepted and rejected by the user with equal importance. We view the user's rejection of particular objects as the end goal in promoting belief harmony.

%Researchers introduce strategies like diversification algorithms and knowledge graph-based recommendations to mitigate the filter bubble. Whereas we introduce the transparent and democratic diversity recommendation model, which acts as an intermediary to spark users' curiosity and guide users affected by filter bubbles. We also emphasize the value of nudge techniques. 

\section{Preliminaries}

In this section, we provide an overview of the fundamental concepts and terminology that are essential for understanding the subsequent discussions and analysis in this paper, which encompasses formal definitions, belief revision, and the category network.

\subsection{Formal Definitions}

A recommendation system is denoted by $S = \langle U, A, C \rangle$, where $U = \{u_{1}, u_{2}, ..., u_{n} \}$ represents a set of $n$ users, $C = \{C_{1}, C_{2}, ..., C_{m} \}$ denotes a set of $m$ categories. $A$ indicates an AI platform designed to recommend a category $C_{i} \in C$ to a user $u_{j} \in U$. 

\vspace{10pt}

\noindent \textbf{Definition 1.} \textit{\textbf{Belief network} of user $u_i$ refers to the representation of $u_i$'s knowledge, beliefs, preferences, and expectations. Mathematically, $u_i$'s belief network is denoted as $G_{u_{i}} = \langle u_i, B_{u_{i}}, C_{u_{i}} \rangle$, where $C_{u_{i}} = \{ C_{u_{i,1}}, C_{u_{i,2}}, ..., C_{u_{i,n}} \}$ represents the set of categories with which the user has previously interacted, and $B_{u_{i}} = \{ B_{u_{i,1}}, B_{u_{i,2}}, ..., B_{u_{i,n}} \}$ signifies the user's belief degree towards each category. In the context of interactions with the BHEISR model, $\Check{M_{u_{i},t}}$ symbolizes the collection of items that have been accepted by $u_{i}$ from time step 0 to t, which includes both system-recommended items and GAI-generated items. Conversely, $\bar{P}{u{i},t}$ denotes the list of GAI-generated prompts that have been presented to $u_{i}$, but whose generated items have been rejected by $u_{i}$ from time step 0 to t.}

\vspace{10pt}

\noindent \textbf{Definition 2.} \textit{\textbf{A category} $C_{i} \in C$ represents a specific category within the recommendation system and is defined as $C_{i} = \langle Sub_{C_{i}}, r_{C_{i}} \rangle$. It consists of a set of subcategories within the category, denoted by $Sub_{C_{i}} = \{Sub_{C_{i,1}}, Sub_{C_{i,2}}, ..., Sub_{C_{i,n}} \}$, and the corresponding click probability for each subcategory, represented by $r_{C_{i}} = \{ r_{C_{i,1}}, r_{C_{i,2}}, ..., r_{C_{i,n}} \}$.}

\vspace{10pt}

\noindent \textbf{Definition 3.} \textit{\textbf{Category Correlation} $\rho(C_{x}, C_{y}) \in [-1, 1]$ refers to the level of relevance between two categories $C_{x}$ and $C_{y}$, with $\rho(C_{x}, C_{y})$ = $\rho(C_{y}, C_{x})$.
A higher value of $\rho(C_{x}, C_{y})$ indicates a stronger correlation between the two categories. To measure category correlation, the title and abstract of all textual items under each category (e.g., news articles or movie descriptions) are processed to extract their features that capture the essential information of the categories. If a piece of item is $N_k$ belongs to the category $C_{x}$, its title and abstract are respectively represented as $t_{N_{k}}$ = $\{w_{1}, w_{2},..., w_{n}\}$ and $a_{N_{k}}$ = $\{a_{1}, a_{2},...,a_{n}\}$, where $n$ means the length of $N_k$'s title and abstract. Further, $N_k$ can be shown as $t_{N_{k}}$ + $a_{N_{k}}$. Then, the extracted features are transformed into numerical representations by employing an embedding representation technique, such as BERT \cite{devlin2018bert}. We transform $N_{k}.t$ to $\vec{N_{k}.t}$ for expressing the representation of $C_{x}$ at time $t$. This transformation allows for capturing the semantic relationships between different categories. The category correlation is formulated as:}

\begin{equation}
\rho(C_{x}, C_{y}) = \frac{C_{x}C_{y}^T}{\left \lVert C_{x} \right \rVert \cdot \left \lVert C_{y} \right \rVert},
\end{equation}

\noindent \textit{where $C_{x}$ and $C_{y}$ denote the vector representations of two categories involved in the calculation.}

Due to the fine-grained feature of BHEISR, we update the category correlation graph after receiving the user feedback. Therefore, if the user accepts the item $N_{g}.{t+1}$ at time $t+1$, the feature of $C_{x}$ at $t+1$ should represent as $\vec{N_{k}.{t+1}}\oplus\vec{N_{g}.{t+1}}$.

\vspace{10pt}

\noindent \textbf{Definition 4.} \textit{\textbf{A Filter Bubble-existed System} $FB^x_{g}$ refers to a system that encompasses a group of \textbf{Filter Bubble-affected Users} $g = \{ u_{n}|u_{n} \in U \}$. In such a system, each user $u_{n} \in g$ exhibits an extreme preference $b^x_{n}$ for a particular category $C_{x}$ while simultaneously being extremely insensitive towards information related to another category.}

\vspace{10pt}

%\noindent $\forall u_{n} \in g$ are significantly influenced by the filter bubble effect in terms of their decision-making behaviors. Specifically, when $FB^x_{g}$ presents an item that involves categories aligned with the user's preference $b^x_{n}$, it is highly likely for the user to accept the item. Conversely, dissimilar items that lack the user's preferred categories face significant challenges in being adopted by $u_{n} \in g$. Thus an item solely consists of a category $C_{y}$ that violates the user's preferences (i.e., $\rho(C_{y}, C_{x}) = -1$) is expected to be rejected by $u_{n} \in g$.

\noindent Each user $u_{n} \in g$ is significantly influenced by the filter bubble effect when it comes to decision-making behaviors. Specifically, when an item presented by $FB^x_{g}$ involves categories that align with the user's preference $b^x_{n}$, the likelihood of the user accepting the item appears high. Conversely, items that do not contain the user's preferred categories face substantial hurdles in gaining the user's acceptance. Thus, an item composed solely of a category $C_{y}$ that contradicts the user's preferences (i.e., $\rho(C_{y}, C_{x}) = -1$) is anticipated to be rejected by $u_{n} \in g$.

%\subsection{Belief Revision}

%Within a user's belief network, both the user and categories are defined as nodes and the user's belief degree is defined as the edges connecting the user node and each category node. The edges signify the strength of the user's preference towards the associated category. In addition, the sum of a user's click probability for all subcategories in the belief network is equal to 1.

%The belief degree is derived from the user's behavior, specifically, their click probability, which indicates their preference in the subcategories. The calculation of belief degree uses entropy as a measurement to quantify the user's click behavior within the subcategories, where a higher value of entropy indicates a stronger belief in the user's preferences within that category.

%Therefore, the calculation of belief degree is based on the user's click probability for all subcategories within the category of interest (i.e., $r_{C_{i}}$). The belief degree $B_{u_{i,j}}$ of user $u_{i}$ towards category $C_{j}$ is formulated as:

\subsection{Belief Revision}

In the context of a user's belief network, both the user and categories are depicted as nodes, with the user's belief degree embodied by the edges connecting the user node to each category node. These edges represent the strength of the user's affinity for the corresponding category. It is important to note that the sum of a user's click probabilities for all subcategories within the belief network must equate to 1.

The belief degree is inferred from the user's behavior, specifically from their click probability, which signifies their inclination towards the subcategories. This click probability is used to calculate belief degree, employing entropy as a metric to quantify the user's click behavior across subcategories. In the current setting, a higher entropy value signifies a more robust belief in the user's preferences for that category.

Thus, the computation of belief degree relies on the user's click probability for all subcategories within the category of interest (denoted by $r_{C_{i}}$). The belief degree $B_{u_{i,C_{j}}}$ of user $u_{i}$ towards category $C_{j}$ is formulated as follows:

\begin{equation}\label{eq:belief}
B_{u_{i,C_{j}}} = - \sum_{k=1}^{N_{Sub_{C_{j}}}} {r_{C_{j,k}}} \log_2({r_{C_{j,k}}}),
\end{equation}

\noindent where $N_{Sub_{C_{j}}}$ signifies the total number of subcategories of $C_{j}$, and $r_{C_{j,k}}$ represents the user's click probability for each individual subcategory under $C_{j}$.

\subsection{Category Network} 

Category correlation quantifies the potential influence between different categories in our recommendation system. When $\rho(C_{x}, C_{y}) = -1$, it signifies a lack of relationship between categories $C_{x}$ and $C_{y}$, indicating that the two categories cannot exert any impact on each other. However, as the value of category correlation approaches 1 (i.e., $\rho(C_{x}, C_{y}) \rightarrow 1$), the correlation between categories $C_{x}$ and $C_{y}$ becomes increasingly strong. This implies that if a user accepts an item from category $C_{x}$, there is a higher probability of the user also accepting items from the highly related category $C_{y}$, and vice versa. In other words, the user's acceptance of items within one category tends to influence their acceptance of items within the correlated category. Therefore, the category network is constructed based on the extent to which the categories are related.

\begin{figure}[htbp]
		\centering
		\includegraphics[width=0.70\textwidth]{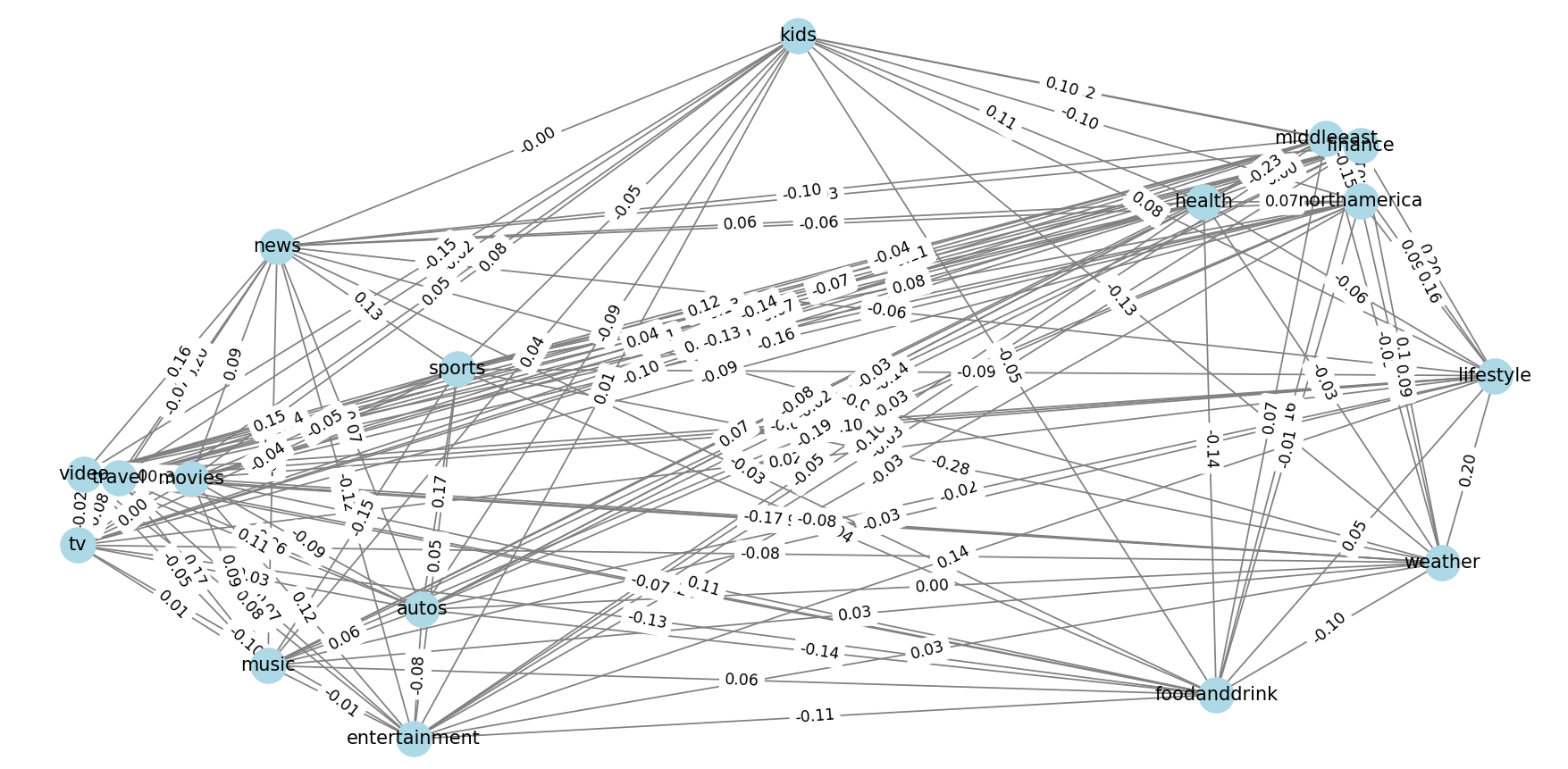}
    \caption{Category correlation graph.}
	 \label{fig:topic_graph}
\end{figure}

The information environment of a recommendation system $S = \langle U, A, C \rangle$ can be represented as a category network. As shown in Figure \ref{fig:topic_graph}, the nodes correspond to the categories $C$ in the system $S = \langle U, A, C \rangle$. Each category $C_{i} \in C$ is considered a distinct node in the category network. The edges connecting these nodes represent the category correlations among the categories, which quantifies the relationships and relevancy between different categories within the system.

% 对于nudge这部分。1.自适应路径探索算法的每个元素，输入输出都是什么。2.对于nudge算法每次自适应路径探索算法的输出输入到nudge算法中，那么这条输入怎么表示。我们的模型是time-sensitive的action logs可能也需要显示出来。3.输入给GAI的信息要怎么表示，输出怎么表示。4.向用户输入的信息怎么表示，用户feedback怎么表示。可以参考yuxuan的论文

\section{The Framework of BHEISR Model}

The BHEISR model is designed to foster a transition for users from selective exposure to belief harmony, effectively enabling them to escape from the filter bubble. The key novelty of BHEISR resides in its incorporation of the "nudge" concept and utilization of GAI for producing recommendation items. This allows users to be gently steered from their highly preferred categories towards less engaging ones, free from any form of compulsion. The proposed BHEISR operates as a mediation mechanism bridging existing recommendation systems and users. It is built based on the existing recommendation algorithm to align with the user's knowledge system but adjusts the user's knowledge system according to the latest round of recommendations.

\begin{figure}[htbp]
		\centering
		\includegraphics[width=0.80\textwidth]{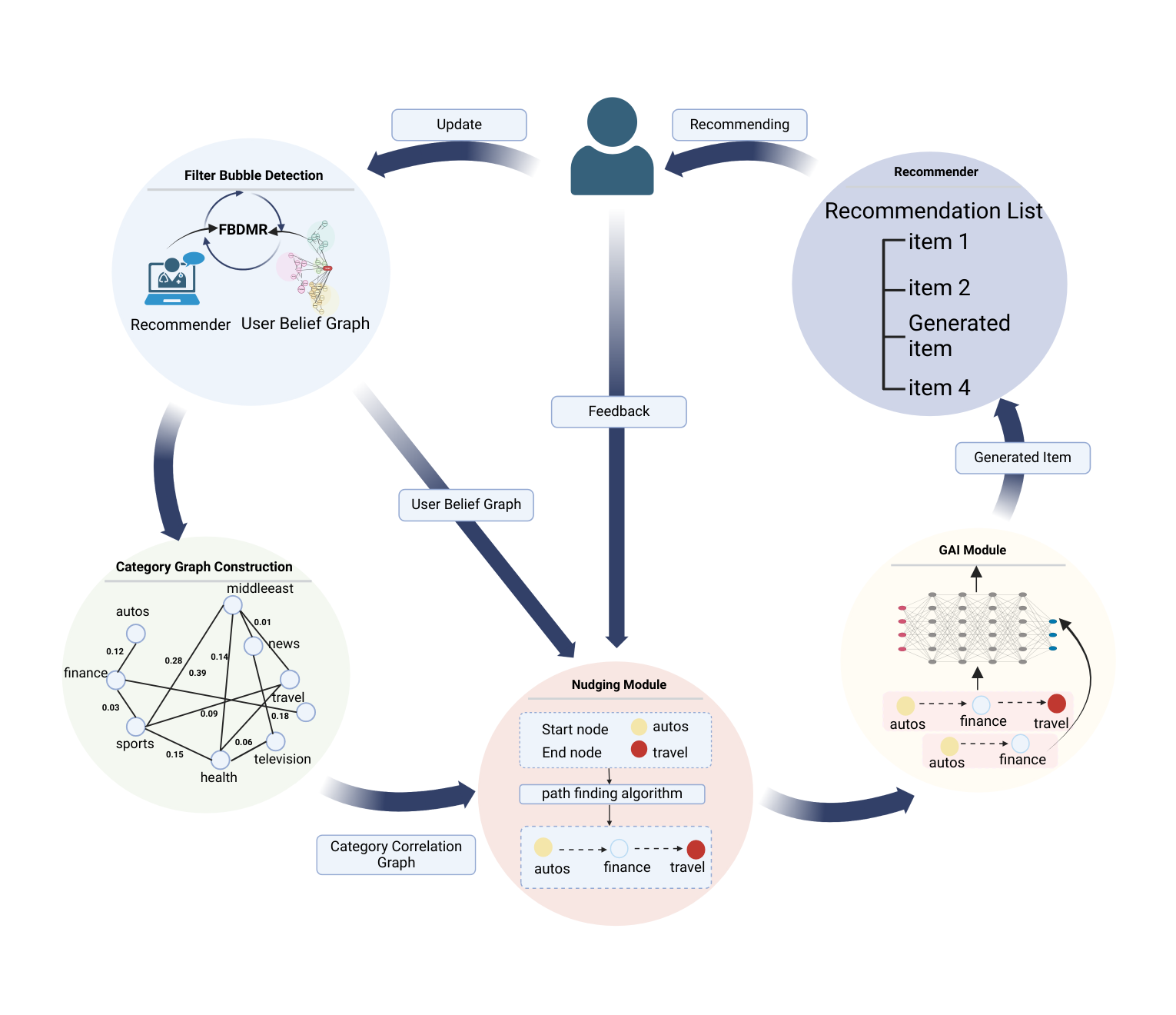}
    \caption{Overall framework of BHEISR model}
	 \label{fig:BHEISR}
\end{figure}

Operating as a dynamic, time-sensitive, and interactive system, the BHEISR model continually adapts users' confidence networks and tailors soft recommendation policies in response to user feedback. In its functioning, it embodies democratic principles and adheres to the principle of non-coercion, thereby fostering subtle and gradual attainment of belief harmony for users. Figure \ref{fig:BHEISR} provides an overview of the entire BHEISR model framework.

%As Figure \ref{fig:BHEISR} represents the whole CHIEISR is a user feedback-aligned time-sensitive recommendation model. The final purpose is to balance users' cognition and avoid filter bubble fatigue and users' selective exposure. The filter bubble detection is the foundation of the BHEISR model. The integration of FR and CR modes in FBDMR offers a comprehensive understanding of filter bubbles, considering system-level biases and user experiences. This holistic approach enables accurate assessment of biased models and supports interventions to mitigate negative impacts.

%In order to achieve category-level nudge recommendations, the fundamental aspect relies on establishing category relationships. The CHIESR model constructs a category relationship graph based on the characteristics of each category and the similarity between pairs of categories. By leveraging the combination of the user's belief graph and topic correlation graph, the model identifies the most optimal paths between categories. Utilizing the GAI framework, the model generates relevant content based on these paths. The final step is to corporate the generated item with existing recommendation lists and present this set of items as a recommendation feed to the user. The model continuously updates users, category networks, and optimal paths based on user interactions, and evaluates the probability of user acceptance until a state of user cognitive harmony is reached.

As demonstrated in Figure \ref{fig:BHEISR}, the BHEISR is a user feedback-responsive, time-sensitive recommendation model. Its objective goal is to balance users' cognition, thereby preventing the exhaustion stemming from the filter bubble phenomenon and users' selective exposure. Detection of the filter bubble forms the foundation of the BHEISR model. The integration of FR and CR modules in FBDMR provides a comprehensive understanding of filter bubbles, taking into account both system-level biases and user experiences. This inclusive approach enables precise evaluation of biased models and facilitates interventions to counteract their adverse effects.

To accomplish category-level nudge recommendations, the essential step is to establish relationships between categories. The BHEISR model builds a category relationship graph, grounded in the unique characteristics of each category and the similarities between category pairs. Leveraging the amalgamation of the user's belief graph and category correlation graph, the model discerns the most suitable paths between categories. By employing the GAI framework, the model generates relevant content based on these paths. The final phase involves integrating this newly generated item with existing recommendation lists and presenting this curated set of items as a recommendation feed to the user. The model continually updates users, category networks, and optimal paths in response to user interactions, while evaluating the probability of user acceptance until the user achieves belief harmony.

\subsection{A Filter Bubbles Detection Model based on Multi-faceted Reasoning (FBDMR)}

%The Filter Bubbles Detection Model based on Multi-faceted Reasoning (FBDMR) is a research model aimed at double-sided investigating existing recommendation systems in terms of their inclination toward user preferences and the existence of selective exposure users. Different from existing solely reasoning models, it operates in both Forward Reconnaissance (FR) and Counter Reconnaissance (CR) modes. The FR analyzes the recommendation system to determine if it exhibits a bias toward user preferences. The CR focuses on users, identifying instances where users are influenced by preference-biased models, resulting in users' conceptual disharmony within filter bubbles. FBDMR provides a comprehensive assessment of filter bubbles, enabling the identification of biased models and the evaluation of their impact on users. It contributes to developing fairer recommendations and interventions to address the adverse effects of filter bubbles. Figure \ref{fig:FBDMR} describes the structure of FBDMR.

\begin{figure}[htbp]
		\centering
		\includegraphics[width=0.80\textwidth]{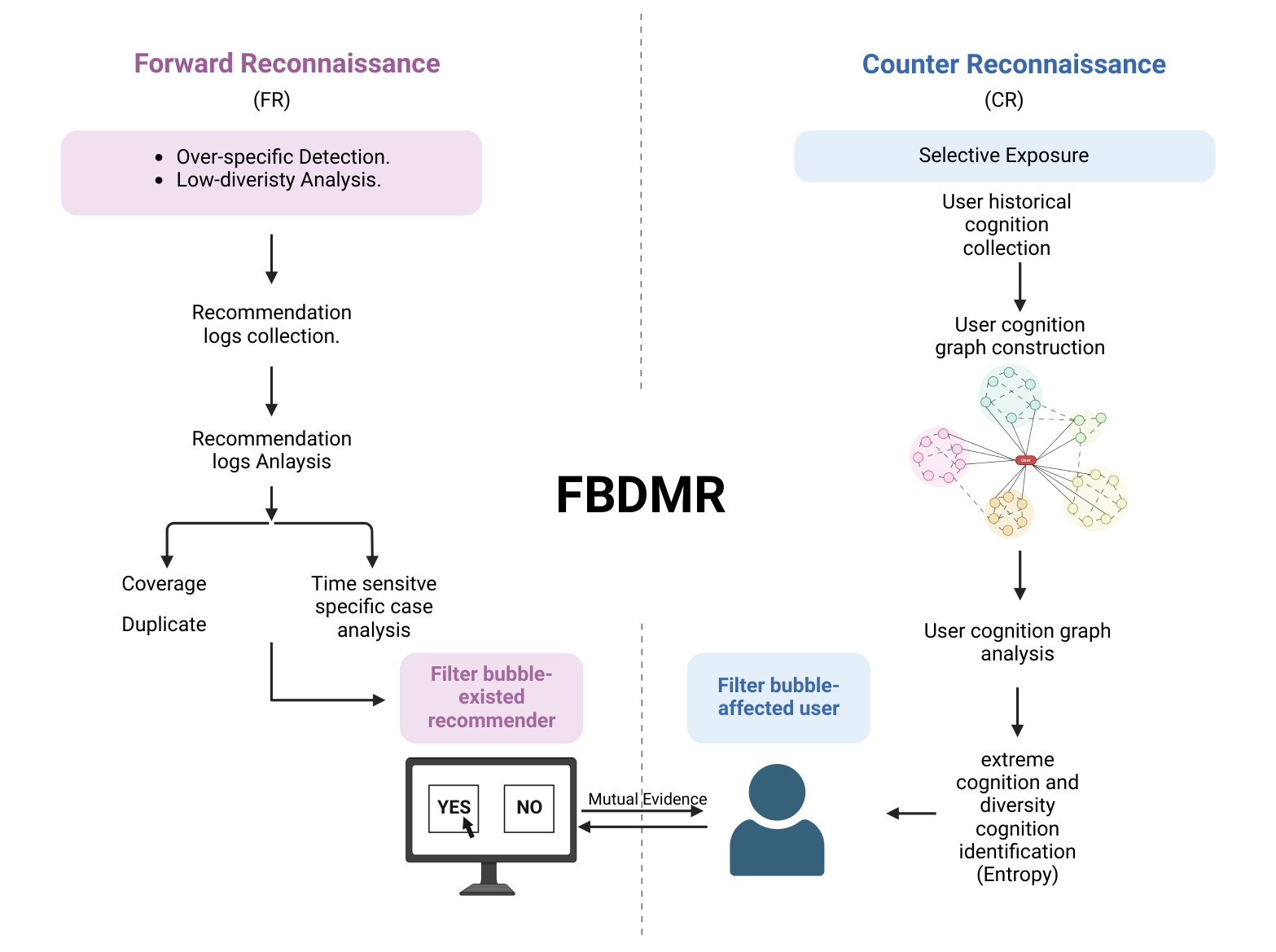}
    \caption{FBDMR: the Filter Bubbles Detection Model based on Multi-faceted Reasoning}
	 \label{fig:FBDMR}
\end{figure}

The Filter Bubbles Detection Model Based on Multi-faceted Reasoning (FBDMR) is a novel model designed to investigate existing recommendation systems from dual perspectives: their inclination towards user preferences and the occurrence of users exhibiting selective exposure. Different from traditional single-dimensional reasoning models, FBDMR operates in both Forward Reconnaissance (FR) and Counter Reconnaissance (CR). The FR assesses the recommendation system to ascertain any potential bias favoring user preferences. In contrast, the CR targets users, pinpointing instances where users are influenced by preference-biased models, which could lead to users' selective exposure within filter bubbles. FBDMR offers an inclusive evaluation of filter bubbles, enabling the detection of biased models and assessing their impact on users. Its implementation contributes to the development of fairer recommendations and interventions to alleviate the negative consequences of filter bubbles. Figure \ref{fig:FBDMR} illustrates the structure of the FBDMR.

\subsubsection{Forward Reconnaissance (FR) Model}
For the Forward Reconnaissance (FR) Model, we validate the user preference bias of the current recommendation model from two perspectives: mathematical validation and time-sensitive data validation. Subcategory coverage score \cite{wang2022user} and subcategory duplicate score \cite{bai2019personalized} are widely used diversity validation metrics, considered as mathematical validations in this paper. The formula for studying the diversity of coverage and duplicate is given below:

\begin{equation}\label{eq:coverage}
DC = \frac{1}{N} \sum_{i=1}^{N} \left(1 - \frac{f_{Sub_{C_{i}}}}{F}\right),
\end{equation}

\noindent where $DC$ indicates diversity coverage in this paper, $N$ means the number of subcategories. $f_{Sub_{C_{i}}}$ is frequency of subcategory $Sub_{C_{i}}$, while $F$ is total frequency of all subcategories.

\begin{equation}
DD = \frac{1}{N(N-1)} \sum_{i=1}^{N} \sum_{j \neq i} d_{ij} ,
\end{equation}

\noindent where $DD$ shows diversity of duplicate in this paper, $N$ means the number of subcategories. $d_{ij}$ represents the duplicate measure between subcategory $Sub_{C_{i}}$ and $Sub_{C_{j}}$.

\begin{figure}[htbp]
		\centering
		\includegraphics[width=0.85\textwidth]{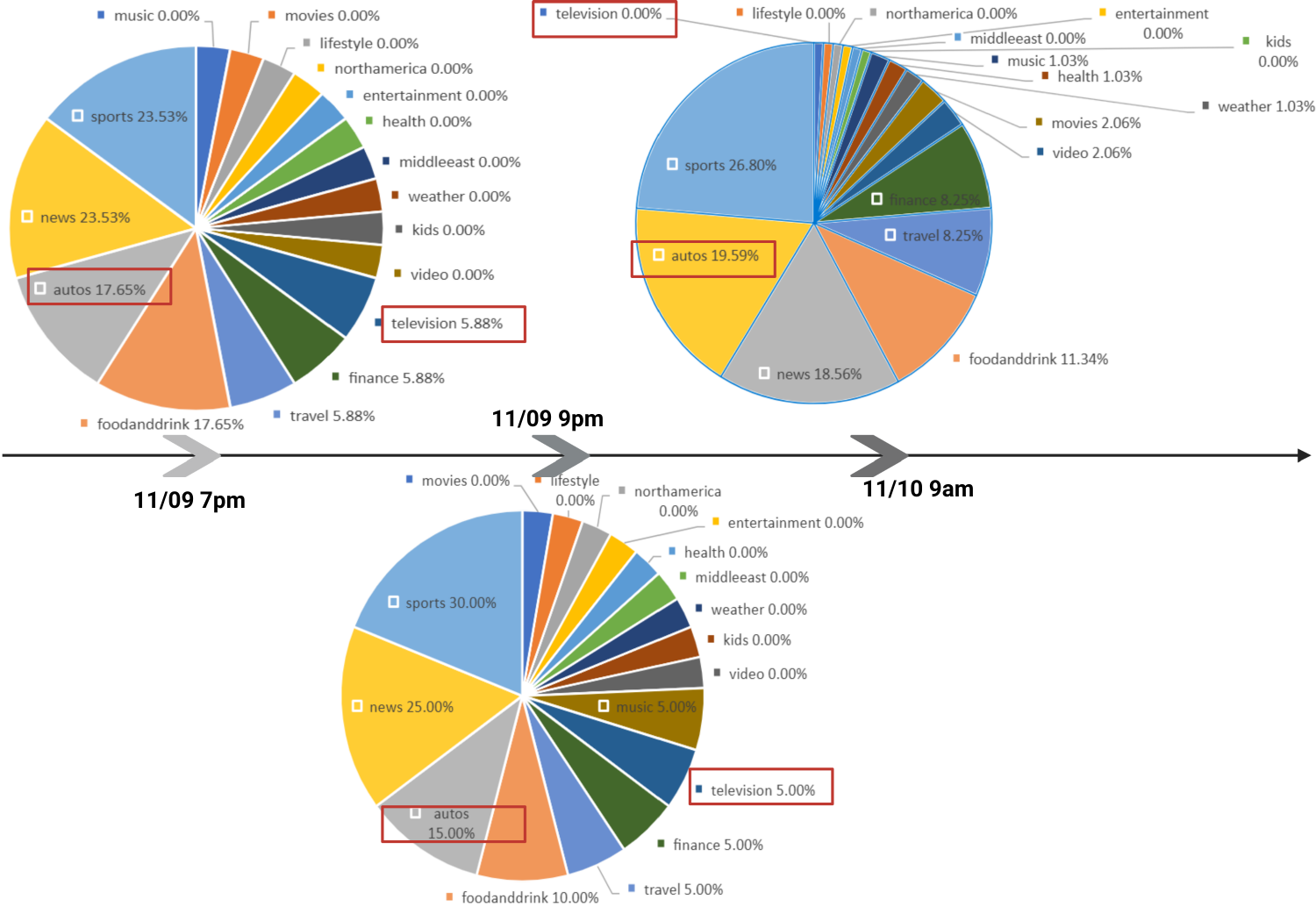}
    \caption{A specific case for filter bubble-existed recommender detection based on time.}
	 \label{fig:time-involving}
\end{figure}

In order to further assess whether the current recommendation algorithm suffers from filter bubbles, we perform an analysis on a specific case by observing the evolution of recommendation logs (existing recommendation items) rankings over time. Figure \ref{fig:time-involving} demonstrates the time-involving recommendation logs.

From the above pie chart, it can be observed that as time progresses, the recommender system consistently recommends information related to the user's preferred categories. In this case, the user shows a greater interest in 'autos' information. Conversely, the system gradually reduces the recommendations for 'television' information as it detects a lack of user interest in that aspect during human-machine interaction. Based on the aforementioned research, it is possible to assess whether the recommender system exhibits a filter bubble. However, further investigation is required to determine whether users are actually being affected by it.

\subsubsection{Counter Reconnaissance (CR) Model}

%The CR Model, centered around the user, is employed to detect the impact of filter bubbles on users. In this study, we construct the user's belief network based on their interaction records with the system. Subsequently, we analyze the user's preferences and disinterest in different categories of information using the belief network. The entropy score is utilized as a differentiating baseline to distinguish between user likes and dislikes. Figure \ref{fig:user_graph} describes a specific user belief network with like and dislike information in our experiment.

The CR model is a user-centric model designed to identify the influence of filter bubbles on user information perception. The model creates a belief network for each user, derived from their interaction records within the system. This belief network is then employed to analyze user preferences and dislikes towards different categories of information.

\begin{figure}[htbp]
		\centering
		\includegraphics[width=0.75\textwidth]{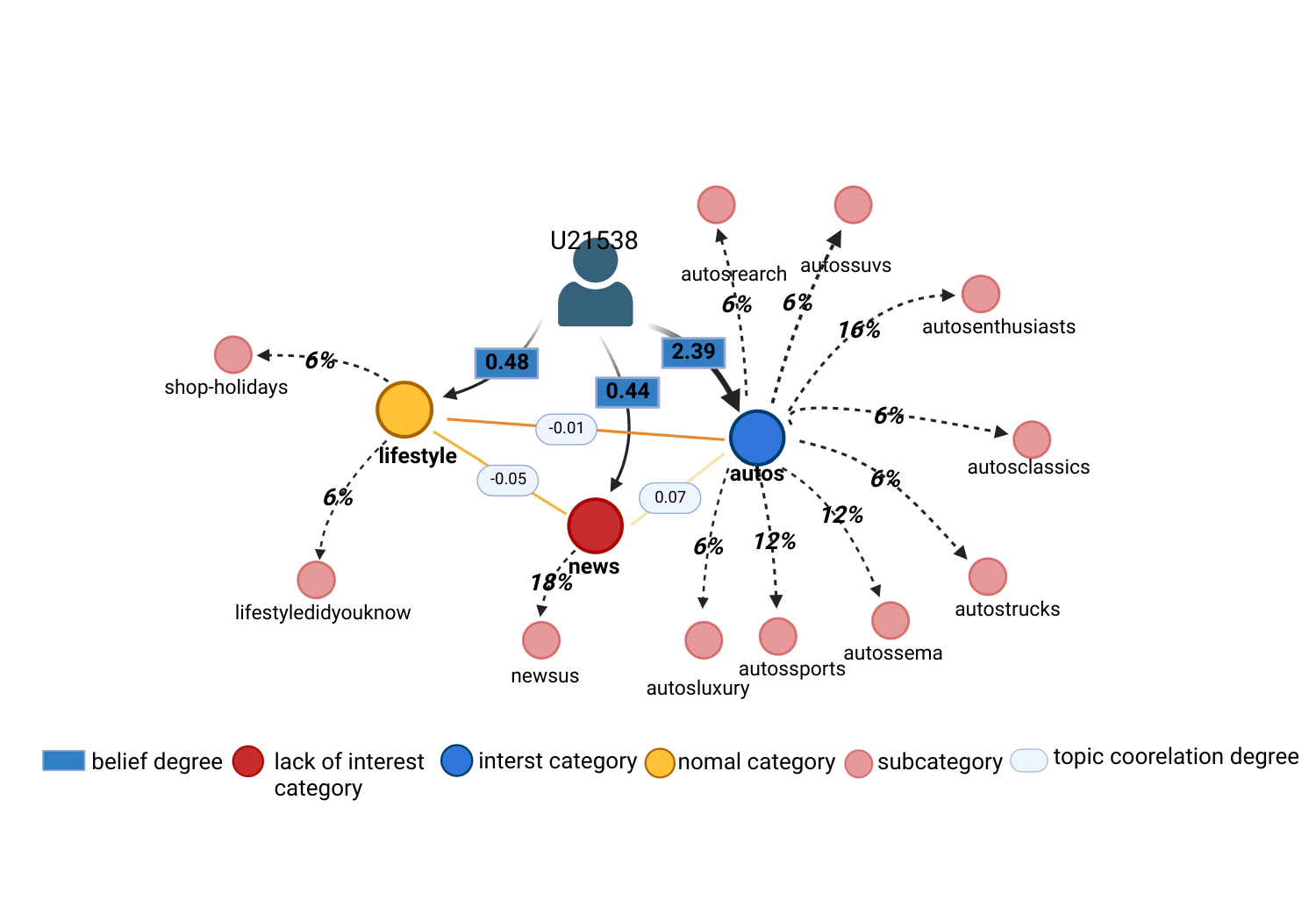}
    \caption{A user-specific belief network incorporating information about preferences and dislikes}
	 \label{fig:user_graph}
\end{figure}

The metric of entropy is employed to establish a baseline for distinguishing user preferences. In particular, high entropy is indicative of diverse user interests, while low entropy suggests a concentration of interest in specific areas. Figure \ref{fig:user_graph} describes a representative user belief network that includes preference and dislikes data used in our experiment.

%The red, blue, and yellow nodes are historical interacted categories $C_{u_{i}}$ of user 'U21538' $u_{i}$, and the pink nodes are subcategories of each category. The edge connecting the category and subcategory indicates the click probability of the user, and the relation connecting the user and category means the user's belief degree $B_{u_{i}}$ towards each category.

%After constructing the users' belief graph, we have given a fresh and explainable definition to determine users with selective exposure. The distribution of the number of people with different cognitive levels is shown in Figure \ref{fig:entropy_distribution} for various types of information.

In Figure \ref{fig:user_graph}, the red, blue, and yellow nodes represent the historically interacted categories $C_{u_{i}}$ of user 'U21538' or $u_{i}$. The pink nodes correspond to the subcategories within each of these primary categories. The edge that links a category to its subcategory represents the user's click probability for that subcategory. Moreover, the connection between the user and a specific category indicates the user's degree of belief towards that category, represented by $B_{u_{i}}$.

\begin{figure}[htbp]
		\centering
		\includegraphics[width=1.0\textwidth]{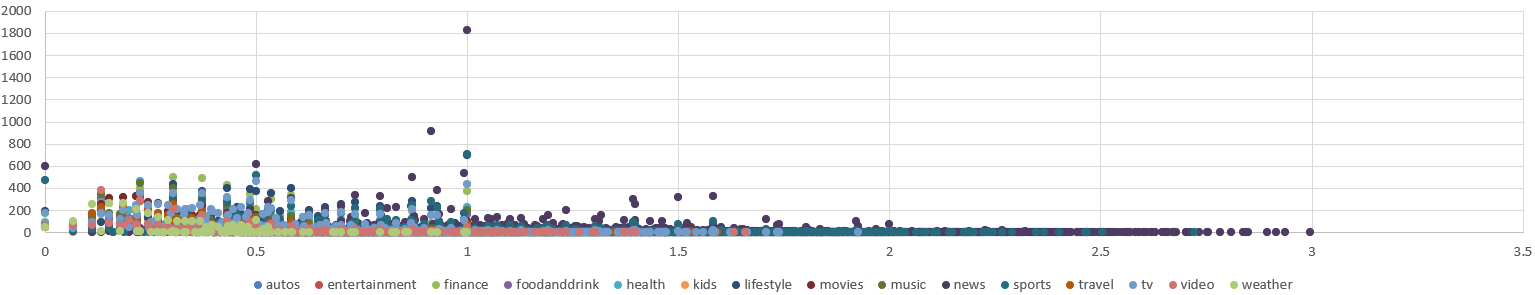}
    \caption{The distribution of individuals with diverse belief levels across various categories}
	 \label{fig:entropy_distribution}
\end{figure}

Upon completion of the user belief graph construction, we introduced a novel and explainable definition for identifying users demonstrating selective exposure to information. The belief diversity among users is presented in Figure \ref{fig:entropy_distribution}, where the distribution of users across various belief levels is depicted for multiple types of information.

We apply the \textit{Kolmogorov-Smirnov} test and \textit{Skewness} calculation to evaluate whether the observed distribution aligns with the standards of a Gaussian distribution, as depicted in the following equations.

\begin{equation}\label{eq:k}
K, p = kstest(N_{C_{m}},'norm', args=(\mu,\sigma)) ,
\end{equation}

\begin{equation}\label{eq:skewness}
skewness_{C_{m}}= \frac{1}{N_{C_{m}}}\sum_{i=1}^{N_{C_{m}}}(\frac{\Tilde{x}-\mu}{\sigma}) ,
\end{equation}

%\noindent where $N_{C_{m}}$ denotes the total number of users in category $C_{m}$. $K$ represents the Kolmogorov-Smirnov statistic, and $p$ represents the p-value of the test. If $p$ is larger than 0.05, it indicates that the distribution of category $C_{m}$ is positively skewed. The skewness of the distribution, denoted as $skewness_{C_{m}}$, can be further analyzed using the \textit{"68-95-99.7"} rule. In equation \ref{eq:skewness}, $skewness_{C_{m}}$ serves as a criterion for classifying skewed distributions in the $m^{th}$ category. In the algorithm, $N_{C_{m}}$ represents the number of users, $\mu$ represents the mean value, and $\sigma$ represents the standard deviation.

\noindent where $N_{C_{m}}$ symbolizes the total number of users within category $C_{m}$. $K$ signifies the Kolmogorov-Smirnov statistic, and $p$ stands for the p-value of the test. A p-value greater than 0.05 suggests that the distribution for category $C_{m}$ is not significantly different from a normal distribution.

The skewness of the distribution, represented as $skewness_{C_{m}}$, offers further insights into the shape of the distribution and can be interpreted in light of the empirical rule (also known as the \textit{"68-95-99.7"} rule). The $skewness_{C_{m}}$ in Equation \ref{eq:skewness} serves as a criterion to classify the degree of skewness in the distributions within the $m^{th}$ category. In this formula, $N_{C_{m}}$ denotes the number of users, $\mu$ represents the mean value, and $\sigma$ represents the standard deviation.

%According to Definition 4, only when a user $u_{i}$ is identified as affected by filter bubbles $u_{i}\in g$ can we determine that the user is experiencing selective exposure, and we can employ our proposed BHEISR approach to achieve cognitive harmony for this user. Facing the positively skewed distribution phenomenon, we conduct the \textit{"68-95-99.7"} rule, to classify users' like and dislike categories.  Figure \ref{fig:entropy} is a specific example for the category 'autos'.

\begin{figure}[htbp]
		\centering
		\includegraphics[width=0.40\textwidth]{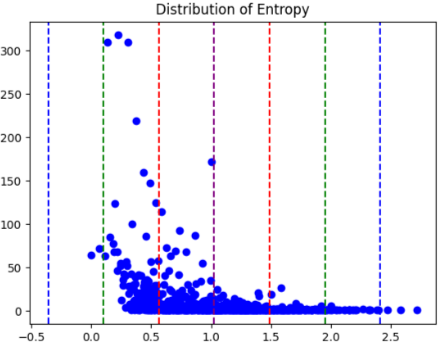}
    \caption{A specific example for the category 'autos'}
	 \label{fig:entropy}
\end{figure}

According to Definition 4, a user $u_{i}$ can be classified as being influenced by filter bubbles, denoted as $u_{i}\in g$, only when they exhibit selective exposure. For distributions exhibiting positive skewness, we apply the empirical rule, or the \textit{"68-95-99.7"} rule, to categorize users' preferences and dislikes. Figure \ref{fig:entropy} provides a specific example, demonstrating this rule's application in the 'autos' category.

%The abscissa is the belief degree of 'autos', and the ordinate is the number of users. According to the "68-95-99.7" rule, $u \in g$ is defined as the user outside the range of two standard deviations of the mean. Here $u \in g$ is defined as the value of belief degree below about 0.1 Users ('autos' is the category the user has no interests) or users whose belief degree is above about 1.9 ('autos' is the category the user has interests).

In Figure \ref{fig:entropy}, the x-axis denotes the belief degree in 'autos', while the y-axis corresponds to the number of users. According to the "68-95-99.7" rule, users outside the range of two standard deviations from the mean are classified as $u \in g$. More specifically, $u \in g$ includes users with belief degrees lower than approximately 0.1 (indicating disinterest in the 'autos' category) and users with belief degrees higher than about 1.9 (indicating strong interest in the 'autos' category).

%Based on the analysis of Figure \ref{fig:user_graph}, it becomes evident that the 'news' category (represented by the red node) exhibits a significantly lower degree of interest, measured at 0.44 belief degree, compared to other categories. Conversely,  this user has extreme interests in the 'autos' category with a 2.39 belief degree. In the context of a user's network, when the user demonstrates a strong interest in one category of information while showing a significant lack of interest in another category, it allows us to classify the user as a cognitively dissonant user influenced by the filter bubble.

As can be observed from Figure \ref{fig:entropy}, it is clear that the user's interests across categories are not evenly distributed. The user demonstrates a significantly lower interest in the 'news' category, denoted by the red node, as indicated by a belief degree of 0.44, compared to other categories. In contrast, the user's interest in the 'autos' category appears exceptionally high, as demonstrated by a belief degree of 2.39. When a user demonstrates a pronounced preference for one category while showing a notable lack of interest in another, this behavior suggests selective exposure influenced by the filter bubble effect. The discrepancy in interest levels allows us to identify and categorize such users accordingly.

\subsubsection{Interplay and Feedback Loop}

%The FR and CR modes are interconnected and form a feedback loop. System biases identified in the FR mode can inform the CR mode's analysis of user experiences, and vice versa. This interplay helps to reveal the complex dynamics between recommendation systems, user preferences, and the resulting filter bubbles. It allows for a more nuanced understanding of how system-level biases can impact individual users and vice versa. In conclusion, the integration of both FR and CR modes in the FBDMR model offers a more holistic and nuanced approach to understanding filter bubbles by considering both system-level biases and user experiences. 

The FR and CR models are linked, forming an intricate feedback loop. Biases identified within the FR model can provide valuable insights for the CR model to better understand user experiences and vice versa. This interaction facilitates a deeper comprehension of the intricate dynamics existing between recommendation systems, user preferences, and the consequent formation of filter bubbles. By examining the interplay between system-level biases and user behavior, we gain insights into how they influence each other. The FBDMR model, combining FR and CR models, offers a comprehensive approach to understanding filter bubbles since it considers system-level biases and user experiences, providing a holistic view of the phenomenon.

%\subsubsection{Summary}
%The proposed FBDMR is a research model that investigates recommendation systems and their alignment with user preferences. Unlike existing models, FBDMR incorporates Forward Reconnaissance (FR) and Counter Reconnaissance (CR) modes to analyze recommendation models and their impact on users. FR examines system bias towards user preferences, while CR focuses on user-level conceptual disharmony caused by preference-biased models. FBDMR offers a comprehensive assessment of filter bubbles, enabling the identification of biased models and the evaluation of their impact. It contributes to developing fairer recommendations and interventions to address the adverse effects of filter bubbles.

\subsection{Nudging Strategy}
%Nudge refers to a behavioral economics approach that aims to induce changes in behavior without relying on financial incentives, aligning with the concept of libertarian paternalism. It involves implementing measures that preserve individuals' freedom of choice while encouraging desired behaviors \cite{jesse2021digital}. The BHEISR model combines knowledge graphs with nudging techniques to provide end-to-end democratic and explainable recommendations. For instance, as depicted in Figure \ref{fig:user_graph}, rather than directly including 'news' in the recommendation list, BHEISR explores a wider range of potential category information to enable nudge recommendations. 

Nudging is a concept drawn from behavioral economics that aims to steer individuals towards certain behaviors without relying on monetary incentives. This concept resonates with the notion of libertarian paternalism, advocating for strategies that preserve individual freedom of choice while subtly guiding towards more beneficial decisions \cite{jesse2021digital}. The BHEISR model leverages knowledge graphs and nudging techniques to deliver democratic and transparent recommendations.

As an illustration, recall that, in Figure \ref{fig:user_graph}, instead of directly incorporating "news" into the recommendation list, the BHEISR model expands the exploration to a broader array of potential category information. This approach serves to facilitate nudging recommendations, subtly guiding users towards diverse information consumption while respecting their freedom of choice.

\subsubsection{Adaptive Path Exploration Algorithms}

%The construction of a recommendation prompt path is the foundation of our nudge recommendation in the BHEISR model. In the BHEISR  model, we adopt an adaptive path exploration algorithm to automatically create the recommendation path on the user feedback for implementing a closed loop for user feedback and system recommendations. The adaptive path exploration algorithm continuously updates itself while perceiving the user and the category graph, in order to recommend prompts that are tailored to the current context. The working process of adaptive path exploration algorithms is shown below. 

The creation of a recommendation prompt path forms the foundation of the nudge recommendation mechanism in the BHEISR model, where an adaptive path exploration algorithm is designed. This algorithm automatically constructs the recommendation path based on user feedback, thereby implementing a closed-loop feedback system that interlinks user feedback and system recommendations.

The adaptive path exploration algorithm is designed to continuously evolve in response to user behavior and changes in the category graph. This dynamic adaptation facilitates the delivery of contextually appropriate prompts, tailored to the user's current situation and preferences. The operational procedure of the adaptive path exploration algorithm is delineated below.

\begin{figure}[htbp]
		\centering
		\includegraphics[width=0.70\textwidth]{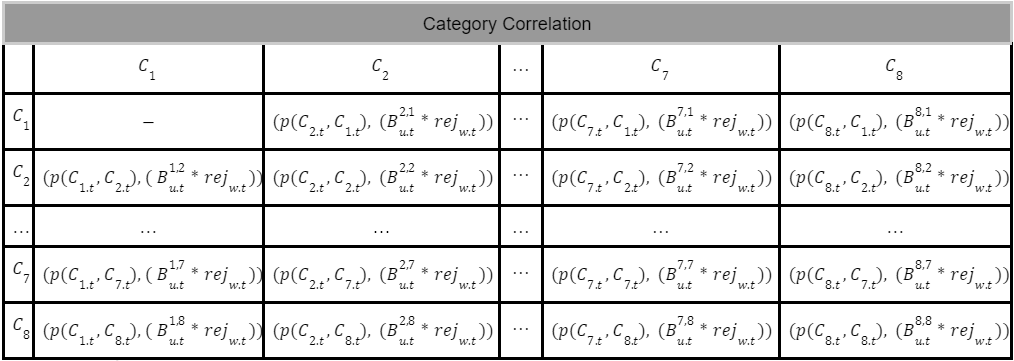}
    \caption{The typical principle of path exploration in the category graph}
	 \label{fig:table_topic}
\end{figure}

The adaptive path exploration algorithm is inspired by the shortest path exploration algorithm, also known as \textit{Dijkstra's algorithm} \cite{barbehenn1998note}. The core principle of this approach is to commence from a central point, traverse the neighboring points, and identify the point with the highest weight as the starting point for the subsequent time step. To accommodate the dynamic nature of user knowledge and category relationships, we have proposed an enhanced adaptive path exploration algorithm. This algorithm integrates the evolving user perceptions and category relationships into the process of path discovery.

%Like Figure \ref{fig:table_topic} shows that there is a category correlation graph $C$ = $\{C_{1}, C_{2},..., C_{i}\}$, $i$ means the number of categories. If $C_{1}$ is the start of a path, the strongest path at the next time stamp should be:

As depicted in Figure \ref{fig:table_topic}, there is a category correlation graph $C = \{C_{1}, C_{2},..., C_{i}\}$, where $i$ denotes the total number of categories. Assuming $C_{1}$ as the start of a path, the most potent path at the next timestamp would be:

\begin{equation}
\rho(C_{1}, C_{n})_{.t+1} = \max(\rho(C_{1}, C_{n})_{.t}+B_{u.t}^{n}*rej_{w.t}),
\end{equation}

\noindent where $C_{n}$ serves as the start node at the timestamp $t+1$, and $B_{u_{i}.t}^{n}$ represents user $u_{i}$'s belief degree of $C_{n}$ at the timestamp $t$. The symbol $w$ signifies the rejection weight, typically set to 1. To ensure the fidelity of user perceptions, we keep the bidirectional relationship graph constant in instances where users decline recommendations, contrasting with the continually evolving belief and categorical relationship networks. We have introduced a tolerance threshold $\theta$ for user rejections. When this tolerance level is exceeded, we assign a rejection weight $rej_{w}$ of -1 to differentiate between disfavored information that changes over time. The optimal prompt path at time $t$, represented as $p_{t}$, is composed of multiple categories. Each category is strongly connected to the starting point at each timestamp until the traversal is completed.

%$C_{n}$ is the start node at the time stamp $t+1$, $B_{u_{i}.t}^{n}$ means user $u_{i}$'s belief degree of $C_{n}$ at time stamp $t$. $w$ represents rejection weight, normally 1. To ensure the accuracy of user perceptions, we maintain the bidirectional relationship graph unchanged in situations where users decline recommendations, compared to the constantly evolving cognitive and categorical relationship networks. We have implemented a tolerance threshold $\theta$ for user refusals, and when the tolerance level exceeds a certain range, we assign a rejection weight $rej_{w}$ of -1 to differentiate uninterested information that changes over time. The optimal prompt path at $t$ can be represented by $p_{t}$ and is constructed from multiple categories, each of which is strongly linked to the starting point at each timestamp until the traversal is completed.

\subsubsection{Nudging and GAI Algorithms}

%We adopt incremental computing function \cite{ramalingam1993categorized} in our nudge process. By decomposing the path into sub-paths, incremental recommendation calculations can be performed. This means that when a user expresses a preference or opinion at a certain position in the path, only the affected sub-path needs to be recalculated instead of recomputing the entire path. It is easier to handle longer or more complex paths compared to simple sequential recommendations, reducing the number of recommendations and increasing efficiency. Figure \ref{fig:nudge_process} is shown below to further describe the recommendation process of BHEISR under the nudge environment. Besides, we provide a detailed description of our nudge strategy in Algorithm \ref{nudge_algorithm}.

\begin{figure}[htbp]
		\centering
		\includegraphics[width=0.80\textwidth]{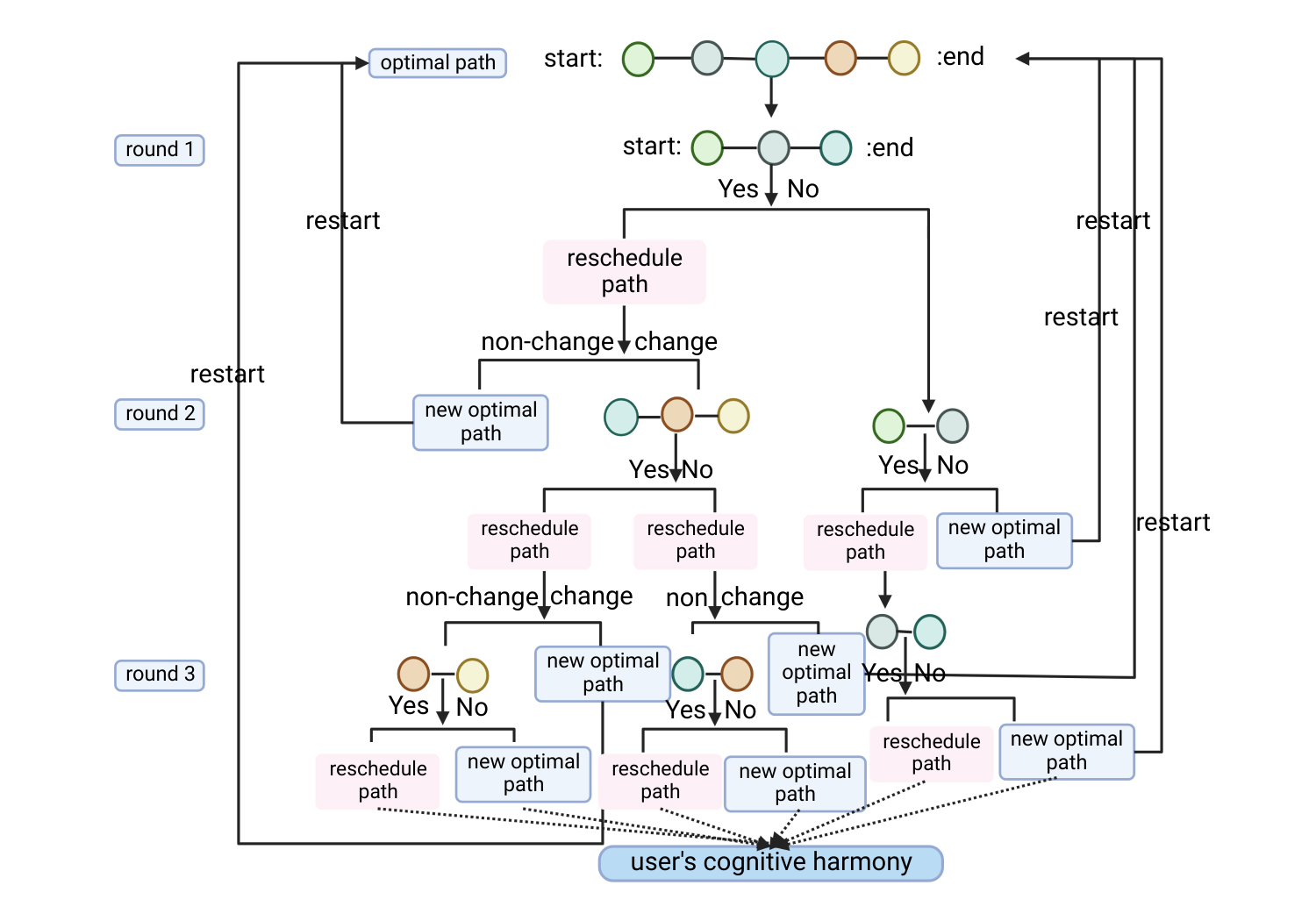}
    \caption{The working process of nudging recommendations}
	 \label{fig:nudge_process}
\end{figure}

The proposed nudging process leverages the concept of incremental computing \cite{ramalingam1993categorized}. By fragmenting the path into smaller sub-paths, we can carry out incremental recommendation calculations. This approach implies that when a user expresses a preference or opinion at a particular point in the path, only the impacted sub-path requires recalibration rather than recomputing the entire path. This method facilitates handling longer or more intricate paths than simple sequential recommendations, thus reducing the number of recommendations and enhancing efficiency. Figure \ref{fig:nudge_process} demonstrates the recommendation process of the BHEISR model within the nudging environment. Furthermore, we describe the proposed nudge strategy in Algorithm \ref{nudge_algorithm}.

\small
\begin{algorithm}
    \caption{Nudge Recommendation Strategy of BHEISR}
    \label{nudge_algorithm}
    \small
    \begin{algorithmic}[1]
        \State Set $p_{t} = \{C_{1}^{p},..., C_{i}^{p}\}$ as the current prompt $p$
        \State Initialise an empty heap $Q=[]$
        \State Q = Binary Split Function(p)
        \While{length(Q) $>0$}
            \State Set current prompt $p=Q[0]$
            \State Generate an item $i_t$ with prompt $p=Q[0]$
            \State Recommend $i_t$ to extreme user $u$
            \If{$u$ accepts $i^1_t$}
                \State $\Check{M}_u = \Check{M}_u \cup i^1_t$ \Comment{$\Check{M}_u$ represents user $u$'s collection of accepted items}
                \State Update user belief graph $B_{u,t}$
                \State Q.pop(0)
            \Else 
                \State $\bar{P}{u,t}.append(p)$ \Comment{$\bar{P}{u,t}$ records user $u$'s declined prompts at time $t$}
                \If{count$(\bar{P}{u,t},p)> \theta$}
                    \State Update category graph
                \EndIf
                \State $p^1_t,p^2_t =$ Binary Split Function (p)
                \State Q.pop(0)
                \State Push $p^1_t,p^2_t$ to Q
            \EndIf
            \If{Binary Split Function (p) is None}
                \State Reschedule path
            \EndIf
        \EndWhile
        \Function{Binary Split Function}{$p_t$}
            \If {$length(p_t) == 2$}
               \State \Return
            \Else
                \If {$mod(length(p_t), 2) == 1$}
                    \State $p^1_t=p_t[0:\frac{length(p_t)-1}{2}]$ 
                    \State $p^2_t=p_t[\frac{length(p_t)+1}{2}:]$
                \Else
                \State $p^1_t=p_t[0:\frac{length(p_t)}{2}]$ 
                    \State $p^2_t=p_t[\frac{length(p_t)}{2}:]$
                \EndIf
            \EndIf
            \State \Return $p^1_t,p^2_t$
        \EndFunction
    \end{algorithmic}
\end{algorithm}
\normalsize

%\textbf{GAI Algorithms} GAI algorithm and nudge recommendation techniques combine effectively in leveraging the interconnectedness of information in a big data environment. This approach overcomes the problem of information gaps in end-to-end recommendation processes. By utilizing the powerful GAI capabilities of \textit{gpt-3.5-turbo}, the method provides rich semantic information for each point in the recommendation path, making the relationships between points more closely linked. It effectively leverages the user's interest in certain information to stimulate their interest in less interesting information.

%As we mentioned, at each timestamp, we can generate a nudging prompt $p_{t}$ = $\{C_{1}^{p},..., C_{i}^{p}\}$. This prompt represents an optimal path between the start node (the most interested category) $C_{1}^{p}$ and the end node (the least interested category) $C_{i}^{p}$.  In order to extract rich information from point-to-point paths in a big data environment, we transmit the paths to GAI to generate context-rich item $GI$ for each point in the path $p_{t}$. The related equation is shown below.

The GAI algorithm, when combined with nudge recommendation techniques, efficiently exploits the interconnectedness of information. This combination presents a solution to address information gaps that may arise during end-to-end recommendation processes. By employing the powerful capabilities of the Large Language Models (LLM), e.g., GPT-3.5 Turbo, this approach offers rich semantic information at each juncture in the recommendation path, thus strengthening the relationships between individual points. This strategy effectively engages a user's interest in specific categories to foster intrigue in areas they might find less attractive.

As previously mentioned, a nudging prompt $p_{t} = \{C_{1}^{p},..., C_{i}^{p}\}$ can be generated at each timestamp. This prompt denotes an optimal path between the starting node, representing the category of highest user interest $C_{1}^{p}$, and the end node, the category of least interest $C_{i}^{p}$. To harness rich contextual information from point-to-point paths within the vast landscape of big data, these paths are fed into the GAI, generating a contextually rich item, $GI$, for each point in the path $p_{t}$. The associated equation is as follows:

\begin{equation}
GI \Rightarrow C_{1}^{p} \cap C_{2}^{p}\cap,...,\cap C_{i}^{p},
\end{equation}

%In summary, the integration of the GAI algorithm and nudge recommendation techniques in the BHEISR model allows for efficient utilization of the interconnectedness of information in a big data environment. It helps overcome information gaps, exploits the rich semantic information provided by GAI, and uses nudge techniques to establish strong connections between information points. The goal is to stimulate user interest in less interesting information based on their interest in other related topics.

In conclusion, the integration of the GAI algorithm and nudge recommendation techniques in the BHEISR model allows the effective utilization of the interconnections. This approach mitigates information gaps, capitalizes on the rich semantic information delivered by the GAI, and employs nudge techniques to establish strong connections between data points. This can help to stimulate user interest in less engaging categories through their established preferences for related subjects.

\subsubsection{Recommendations}

%As an intermediary between the recommendation system and users, the BHEISR model alleviates the negative impact of filter bubbles caused by the current system while promoting cognitive balance for users experiencing selective exposure. We add an additional set of lists to the original recommendation list, where the final recommended user list is represented as $feed$, such that $feed$ = $\{feed_{orignal}, feed_{current}\}$. $feed_{original}$ is consisted of several items $i$, which represents as $feed_{original}$ = $\{i_{1}, i_{2},..,i_{j}\}$. To obtain the original recommendation list, two filtering algorithms are employed: Content-Based filtering (CB) \cite{reddy2019content} and User-Collaborative filtering (UC) \cite{wang2021robust}. 

%CB estimates the utility of an item $i$ to a user $u$ based on the content of the item. This utility is computed by the similarity between the content of the item and the user's historical accepted items:

As an intermediary between the recommendation system and users, the BHEISR model alleviates the adverse effects of filter bubbles caused by the current system and fosters belief balance among users subjected to selective exposure. We introduce an additional set of lists to the original recommendation list, where the final user recommendation list is denoted as $feed$, such that $feed$ = $\{feed_{original}, feed_{current}\}$. $feed_{original}$ consists of several items $i$, represented as $feed_{original}$ = $\{i_{1}, i_{2},..,i_{j}\}$. To derive the original recommendation list, we employ two filtering algorithms: Content-Based Filtering (CB) \cite{reddy2019content} and User-Collaborative Filtering (UC) \cite{wang2021robust}.

CB determines the utility of an item $i$ for a user $u$ based on the item's content. This utility is computed by assessing the similarity between the content of the item and the historical items accepted by the user:

\begin{equation}
    \label{eq:cb}
    score_{i,u} = sim(TF-IDF_i, \frac{1}{n} \sum_{j=1}^{n} \text{{TF-IDF}}_j),
\end{equation}

%\noindent where $sim(\cdot)$ computes the cosine similarity between an item $i$ and a user $u$. TF-IDF is used to analyze the content of the item $i$. $\frac{1}{n} \sum_{j=1 | j \in M_{u,t}}^{n} \text{{TF-IDF}}_j$ describes user's preference according to its historical behaviour. To be specific, $\Check{M}_{u,t}$ is the user's historical records at time step $t$, which suggests the items it clicked from time 0 to t. $\frac{1}{n} \sum_{j=1 | j \in M_{u,t}}^{n} \text{{TF-IDF}}_j$ is the average of TF-IDF feature vectors of all interacted items. UC considers the utility of an item $i$ in the historical records of other similar users in Equation \ref{eq:uc}. The similarity between two users is calculated using cosine similarity, taking into account their respective historical records. Additionally, $B_{u_{i}}$ represents the vector of user $u_i$ with respect to different categories.

\noindent where $sim(\cdot)$ calculates the cosine similarity between an item $i$ and a user $u$. We use TF-IDF to analyze the content of the item $i$. $\frac{1}{n} \sum_{j=1 | j \in M_{u,t}}^{n} \text{{TF-IDF}}j$ represents the user's preferences according to their historical behavior. Specifically, $\Check{M}{u,t}$ signifies the user's historical records at time step $t$, indicating the items they clicked from time 0 to t. $\frac{1}{n} \sum_{j=1 | j \in M_{u,t}}^{n} \text{{TF-IDF}}j$ is the average of TF-IDF feature vectors of all interacted items. UC assesses the utility of an item $i$ based on the historical records of other similar users as per Equation \ref{eq:uc}. We calculate the similarity between two users using cosine similarity, considering their respective historical records. Moreover, $B{u_{i}}$ denotes the vector of user $u_i$ across different categories.

\begin{equation}
    \label{eq:uc}
    score_{i \in \Check{M}_{j},u} = \frac{\Check{M}_{i}\Check{M}_{j}^T}{\left \lVert \Check{M}_{i} \right \rVert \cdot \left \lVert \Check{M}_{j} \right \rVert} * B_{u_{i}}
\end{equation}

$feed_{current}$ indicates $GI$, so the recommended feed to users should be $feed$ = ${i_{1}, i_{2},..,i_{j}, GI}$. After the BHEISR model recommends a $feed$ to the target filter bubble-affected user $u$, $u$ decides whether to accept the recommendation based on the acceptance probability $AP^i_{u,t}$, as defined in Equation \ref{accept probability}:

\begin{equation}
    \label{accept probability}
    AP^{i_j}_{u,t} =\sum_{C_x \in i_j} \omega_{C_x} * \frac{B_{u,n}}{\sum_{i=1}^{n} B_{u}},
\end{equation}

\noindent where $\omega_{C_x}$ is the weight of category $C_x$ contained within the content of an item $i_j \in feed$. $B_{u_i,n}$ represents the belief degree of $u$ towards category $C_i$ at the current time step. $\sum(B_{u,t})$ calculates the total belief degrees of $B_{u,n}$.

\section{Experiments}

We conduct experiments from two general directions: system and user. Regarding the system-centered experiment, it is mainly to prove the effectiveness of the BHEISR model as an intermediate agency in alleviating the system filter bubble. Regarding the user’s perspective, we conducted four user-centered experiments, including detecting the positive effect of the BHEISR model on increasing user belief diversity, examining the effectiveness of the BHEISR model in motivating filter bubble-impacted users to the category they are less interested in, and analyzing the ability of the BHEISR model to reduce the number of filter bubble-impacted users. Finally, we conducted a parametric analysis experiment to analyze the different effects of different BHEISR recommendation weights on stimulating users' interest in new categories.

\subsection{Experimental Setting}

\subsubsection{\textbf{Dataset}}

%This section describes the experimental setup for evaluating the performance of the proposed BHEISR. 
In the experiments, we leverage two real-world datasets from two sources: the Microsoft News dataset (MIND) \footnote{https://msnews.github.io/} and IMDB dataset \footnote{https://www.kaggle.com/datasets/meastanmay/imdb-dataset?select=tmdb\_5000\_movies.csv/}. \\

\noindent \textbf{MIND Dataset} is a large-scale and publicly available news recommendation dataset consisting of user interaction data collected from Microsoft News. The dataset contains 50,000 samples, and a smaller, lighter version includes 5,000 users with 230,117 user behaviors. Each behavior captures a user's reading habits and includes information such as user ID, timestamp, category, subcategory, title, and click behavior. In the MIND dataset, click behavior acts as a binary indicator of user interest, where a value of 1 signifies that the user reads the article, while a value of 0 indicates dislike.\\

\noindent \textbf{IMDB Dataset} includes movie rating records that reflect users' past behaviors. The dataset provides movie-related information, including movie ID, genres, titles, and overviews, along with user ratings for movies and corresponding user IDs. In our experiments, the degree of a movie's rating is considered a proxy for user interest, with ratings ranging from 0 to 5. A rating higher than 2.5 indicates user interest in the movie, whereas a rating below that suggests a lack of interest.\\

\noindent \textbf{Datasets Statistics}. Table \ref{table:datasets} provides an overview of the sizes and properties of the two datasets, IMDB and MIND. It shows statistics on the number of users, user behaviors, categories, and filter bubble-affected users. The category diversity is similar across the two datasets, with 16 and 17 categories each. The users who exhibited extreme belief imbalance are selected for the experiments. In the MIND dataset, we identified 180 such users, and from the IMDB dataset, we found 20 out of 300 individuals that met the criteria. The experiments focus on this select group of 200 users.

\begin{table}[t]
\centering
\begin{tabular}{l|ll}
\hline
\textbf{Dataset} & \textbf{IMDB} & \textbf{MIND} \\ \hline
\textbf{No. users} & 333  & 5,000\\
\textbf{No. behaviours} & 105,340 & 230,117\\
\textbf{No. categories} & 16 & 17\\
\textbf{No. $FB_{affected}$ users}   & 20 & 180\\\hline
\end{tabular}%
\caption{Statistics of the datasets}
\label{table:datasets}
\end{table}

\textbf{Evaluation of User Feedback.} Given the inherent impracticality and high cost associated with online testing for researchers, we have designed a novel offline evaluation approach:

\begin{enumerate}[label={[\arabic*]}]
    \item Implement an 'Acceptance Probability Algorithm' to simulate user feedback.
    \item Generate recommendations using a 'Nudge Strategy' based on the simulated user feedback.
    \item Evaluate the recommendations by considering aspects such as diversity, and efficacy in mitigating filter bubbles. This includes employing metrics like Category Coverage and User Belief-related algorithms.
\end{enumerate}

\subsection{Parameter settings and baselines}

\textbf{Baselines:} We assess the performance of BHEISR in comparison with several established baseline methods:

\begin{itemize}
    \item Random (RD): This method suggests a selection of existing items randomly.
    \item Content-Based Filtering (CB): This strategy recommends existing items based solely on content-based filtering.
    \item User-Collaborative Filtering (UC): This approach recommends existing items using only user collaborative filtering.
    \item BHEISR: This method suggests a set of items (designated as $GI$) as recommendation feeds, derived using the BHEISR approach.
    \item Random with BHEISR ($RD_{w}C$): This model proposes a list of recommended items, comprising randomly selected existing items and items derived from BHEISR.
    \item CB with BHEISR ($CB_{w}C$): This method offers a list of recommended items that combine items selected through content-based filtering and items suggested by BHEISR.
    \item UC with BHEISR ($UC_{w}C$): This approach generates a list of recommended items, including items selected by user collaborative filtering and items recommended by BHEISR.
\end{itemize}

\textbf{Evaluation Metrics:}
We evaluate the BHEISR model from both system and user perspectives. From a system perspective, we use the Coverage degree to judge the diversity of system recommendations. The \textbf{Coverage algorithm} refers to Section 4.4.1, where the coverage degree algorithm is formulated in Equation \ref{eq:coverage}. 
From the users' perspective, we conduct experiments in three aspects: detecting the diversity of users' belief networks, detecting users' interests in categories that users are not interested in, and the number of users affected by filter bubbles. We still use the Coverage algorithm to judge the diversity of user belief networks. 
The \textbf{user belief algorithm} described in Equation \ref{eq:belief} of Section 3.2. Equations \ref{eq:skewness} and \ref{eq:k}, and the previously mentioned '68-95-99.7' rule in Section 4.1.2 are the metrics to check filter bubble-impacted users and user belief degrees for different categories. Besides, we also make the parameter analysis for detecting the user belief changes on different nudge weights. The metric in this experiment also utilizes the user belief algorithm.

\textbf{Parameters:} 
The proportion of BHEISR-generated items ($w$) is seen as a parameter in our experiments. We analyze the impact of different proportions $w$ within a recommendation $feed$ on user belief diversity and recommendation system diversity.

\subsection{Experimental Results}

\subsubsection{Experiment 1: \textbf{Coverage Analysis}}

The first experiment analyzes the impact of the proposed BHEISR on both the diversity of the baselines and the user belief network. We randomly select a filter bubble-affected user from both datasets as our experimental object in this case. 'U25354' is chosen from the MIND dataset, while 'U128' is selected from the IMDB dataset.

\textbf{Coverage Analysis for systems.} First of all, we examine the evolution of diversity in recommendation lists over time for seven models, each based on filter bubble-affected users from different domains. The results are demonstrated in Tables \ref{tab:coverage_MIND} and \ref{tab:coverage_IMDB}. Here, $"\textit{sum.}"$ signifies the total coverage degree of each model throughout the recommendation process, and $"\textit{Improv.}"$ denotes the rate of growth in diversity.

\tiny
\begin{table}[htbp]
\centering
\renewcommand{\arraystretch}{0.60} % Increase row height
\resizebox{\textwidth}{!}{
\begin{tabular}{@{}l|lllllll@{}}
\hline
\multicolumn{8}{c}{\textbf{MIND}} \\ \hline
\textbf{Times} & \textbf{RD} & \textbf{$RD_{w}C$} & \textbf{CB} & \textbf{$CB_{w}C$} & \textbf{UC}  & \textbf{$UC_{w}C$} & \textbf{BHEISR} \\ \hline
$feed_{1}$ & 0.176 &  0.411 & 0.058 & 0.294 & 0.152 & 0.294 & 0.294\\\hline
$feed_{2}$ & 0.235 & 0.411 & 0.058 & 0.117 & 0.235  & 0.235 & 0.117 \\\hline
$feed_{3}$ & 0.176 & 0.294& 0.058 & 0.117 & 0.176 & 0.235 & 0.117 \\\hline
$feed_{4}$ & 0.117 & 0.294 & 0.058 & 0.176 & 0.117 & 0.294 & 0.117 \\\hline
$feed_{5}$ & 0.176 & 0.294 & 0.058 & 0.117 & 0.176 & 0.235 & 0.117 \\\hline
$feed_{6}$ & 0.235 & 0.294 & 0.058 & 0.235 & 0.117 & 0.235 & 0.117 \\\hline
$feed_{7}$ & 0.294 & 0.235 & 0.058  & 0.117 & 0.235 & 0.294 & 0.235 \\\hline
$feed_{8}$ & 0.235 & 0.235 & 0.058 & 0.176 & 0.294 & 0.294 & 0.117  \\\hline
$feed_{9}$ & 0.235 & 0.176 & 0.058 & 0.235 & 0.176 & 0.235 & 0.117  \\\hline
$feed_{10}$ & 0.235 & 0.294 & 0.058 & 0.176 & 0.294 & 0.176 & 0.352  \\\hline
$\textit{sum.}$ & 2.114 & \textbf{3.232} & 0.580 & 1.760 & 1.972  & \textbf{2.527} & 1.695  \\\hline
$\textit{Improv.}$  & $RD_{w}C$ - $\textbf{RD}$ & $\textbf{52}\%$   & $CB_{w}C$ - $\textbf{CB}$ & $\textbf{203.45}\%$ & $UC_{w}C$ - $\textbf{UC}$ & $\textbf{28.14}\%$\\\hline
\end{tabular}
}
\caption{Coverage Analysis of Recommendation Strategies based on MIND datasets}
\label{tab:coverage_MIND}%
\end{table}
\normalsize

\tiny
\begin{table}[htbp]
\centering
\renewcommand{\arraystretch}{0.5} % Increase row height
\resizebox{\textwidth}{!}{
\begin{tabular}{@{}l|lllllll@{}}
\hline
\multicolumn{8}{c}{\textbf{IMDB}} \\ \hline
\textbf{Times} & \textbf{RD} & \textbf{$RD_{w}C$} & \textbf{CB} & \textbf{$CB_{w}C$} & \textbf{UC} & \textbf{$UC_{w}C$} & \textbf{BHEISR}  \\ \hline
$feed_{1}$ & 0.125 & 0.312 & 0.062 & 0.312 & 0.312 & 0.500 & 0.500  \\\hline
$feed_{2}$ & 0.187 & 0.250 & 0.062 & 0.125 & 0.250 & 0.370 & 0.125  \\\hline
$feed_{3}$ & 0.125 & 0.125 & 0.062 & 0.125 & 0.187 & 0.250 & 0.312  \\\hline
$feed_{4}$ & 0.187 & 0.187 & 0.062 & 0.125 & 0.312 & 0.250 & 0.187  \\\hline
$feed_{5}$ & 0.250 & 0.125 & 0.062 & 0.125 & 0.250 & 0.187 & 0.125   \\\hline
$feed_{6}$ & 0.250 & 0.250 & 0.062 & 0.187 & 0.250 & 0.250 & 0.125   \\\hline
$feed_{7}$ & 0.125 & 0.187 & 0.062 & 0.25 & 0.250 & 0.125 & 0.187  \\\hline
$feed_{8}$ & 0.187 & 0.25 & 0.062 & 0.187 & 0.250 & 0.312 & 0.187   \\\hline
$feed_{9}$ & 0.187 & 0.187 & 0.062 & 0.312 & 0.312 & 0.187 & 0.187   \\\hline
$feed_{10}$ & 0.250 & 0.187 & 0.062 & 0.125 & 0.187 & 0.250 & 0.125   \\\hline
$\textit{sum.}$ & 1.873 & \textbf{2.06} & 0.620 & 1.873 & 2.56 & \textbf{2.681} & 2.06  \\\hline
$\textit{Improv.}$  & $RD_{w}C$ -$\textbf{RD}$ & $\textbf{9.983\%}$   & $CB_{w}C$ - $\textbf{CB}$& $\textbf{202.09\%}$ & $UC_{w}C$ - $\textbf{UC}$ & $\textbf{21.6\%}$\\\hline
\end{tabular}
}
\caption{Coverage Analysis of Recommendation Strategies based on IMDB datasets}
\label{tab:coverage_IMDB}%
\end{table}
\normalsize

%Comparing both Table \ref{tab:coverage_MIND} and Table \ref{tab:coverage_IMDB}, the same model will perform roughly the same on different users. Random, UC, and CB have relatively low coverage rates, and their values show minimal variation over time. This suggests that these strategies may be preference-based recommendation models, lacking significant advantages in terms of coverage range or variability. The strategies $CB_{w}C$ and $UC_{w}C$ exhibit higher coverage rates, and their values show significant increases over time. This indicates that these strategies perform relatively well in terms of coverage and demonstrate an expanding coverage range as time progresses. The strategy $RD_{w}C$ shows a higher coverage rate than among all strategies. However, its coverage values remain relatively stable over time, without significant increases or variations.

Upon comparison of Table \ref{tab:coverage_MIND} and Table \ref{tab:coverage_IMDB}, it can be observed that the performance of the same model across different users remains constant. The strategies Random, UC, and CB exhibit relatively lower coverage rates, with their values demonstrating minor fluctuations over time. This pattern suggests that these strategies may represent preference-based recommendation models, which lack coverage breadth or variability advantages. In contrast, the strategies $CB_{w}C$ and $UC_{w}C$ show higher coverage rates, with a significant increase in these values over time. This observation points to the relative efficacy of these strategies in coverage, as well as their progressively expanding coverage range. While the strategy $RD_{w}C$ exhibits the highest coverage rate among all strategies, its coverage values remain fairly consistent over time without notable increases or variations.

%It can be seen from $\textit{Improv.}$ that all BHEISR-based recommendation systems exhibit wider diversity in recommendations. $RD_{w}C$ and $UC_{w}C$ strategies outperform the other strategies in terms of coverage, with $UC_{w}C$ showing the most significant improvements over time. $RD_{w}C$ strategy also demonstrates a relatively higher coverage rate but lacks the increasing trend observed in the aforementioned strategies. Besides, as an independent recommendation model, BHEISR is not so prominent in breaking the influence of the system filter bubble.

The values of $\textit{Improv.}$ reveal that all BHEISR-based recommendation systems present greater diversity in recommendations. The strategies $RD_{w}C$ and $UC_{w}C$ outperform the rest in terms of coverage, with $UC_{w}C$ demonstrating the most substantial improvements over time. While $RD_{w}C$ strategy displays a comparably higher coverage rate, it lacks the upward trend observed in the previously mentioned strategies. Furthermore, when considered as an independent recommendation model, BHEISR does not appear to be particularly effective at breaking the influence of the system filter bubble.

\begin{figure}[htbp]
\centering
\subfigure[Coverage analysis of user beliefs on the MIND datasets]{\includegraphics[width=0.450\textwidth]{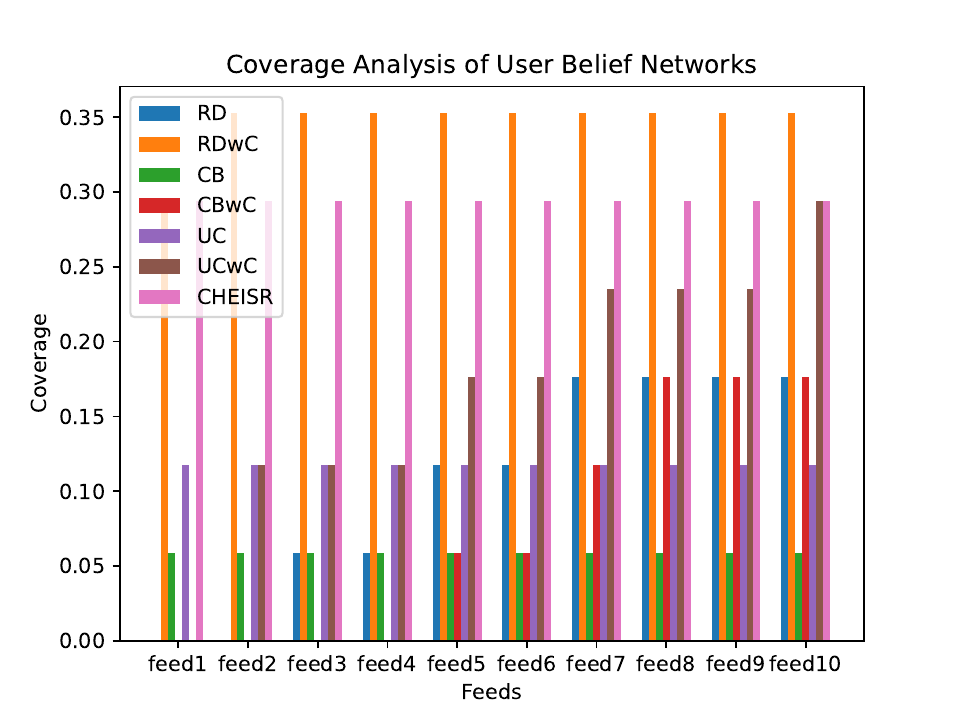}\label{fig:coverage_user_belief}}
\subfigure[Coverage analysis of user beliefs on the IMDB datasets.]{\includegraphics[width=0.450\textwidth]{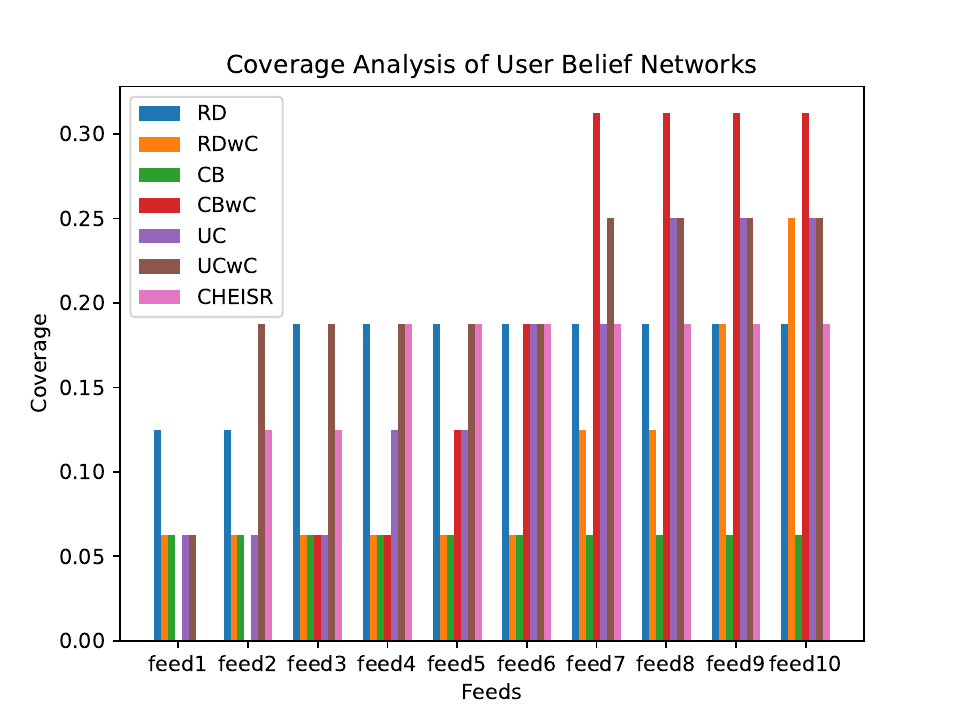}}\label{fig:coverage_user_belief_IMDB}
\caption{Coverage analysis of user beliefs on diverse datasets}
\label{fig:parameter}
\end{figure}

\textbf{Coverage Analysis for User Belief Networks.} After analyzing the effectiveness of the BHEISR approach in mitigating the filter bubble.  we conduct another experiment to investigate its efficiency in enhancing the diversity of user belief across multiple domains. Figure \ref{fig:parameter} demonstrates the dynamic changes in belief diversity for time-sensitive users' beliefs.

%From these two bar charts in Figure \ref{fig:parameter}, there may be some differences in the details. This could be due to variations in the datasets. However, overall, the general trends are consistent in both bar charts.

The two bar charts in Figure \ref{fig:parameter}, though bearing potential discrepancies in details, attributable to variations within the datasets, generally align in their overarching trends.

In the comparison of RD and $RD_{w}C$, the $RD_{w}C$ model exhibits superior performance. For the users in the MIND dataset, their openness to new information remains high and stable. Conversely, for the user in the IMDB dataset, the model's expressiveness exhibits a growing trend. As the number of recommendations escalates, users begin to slowly accept more new information. The results suggest that the combination of random selection and BHEISR strategies can help to increase the diversity within users' belief systems. On top of that, comparisons between UC and $UC_{w}C$, as well as CB and $CB_{w}C$, indicate that the $UC_{w}C$ and $CB_{w}C$ models surpass the UC and CB models, respectively. This indicates that the integration of the BHEISR strategy improves the performance of the UC and CB models. 

%Overall, the models incorporating the BHEISR strategy ($RD_{w}C$, UCwBHEISR, $CB_{w}C$) perform better in their respective comparisons compared to the models without the BHEISR strategy (RD, UC, CB). As the number of recommendations increases, the BHEISR-based recommendation model leads to greater diversity in user beliefs. This indicates that users are progressively developing an increased interest in obtaining more varied information over time.

Overall, the models incorporating the BHEISR strategy ($RD_{w}C$, $UC_{w}C$, $CB_{w}C$) outperform their counterparts lacking the BHEISR strategy (RD, UC, CB) in their respective comparisons. With an increasing number of recommendations, the BHEISR-based recommendation model fosters a broader diversity in user beliefs. This reveals a growing tendency among users to seek more diverse information over time.

\subsubsection{Experiment 2: \textbf{User Beliefs Analysis}}

%The BHEISR model is built upon existing recommendation systems to deliver information to users. Its primary objective is to achieve a balance in the user's belief network. Users' cognitive changes require long-term guidance, so in order to demonstrate that users are accepting information they are less interested in through recommendations in a relatively short time period, we selected users with shorter recommendation paths $p_{\textit{shorter}.u}$ (users whose interested knowledge and less interested knowledge have closer feature distances) from the IMDB dataset, user 'U276'. This approach aims to show changes in user beliefs in a short period of time. Besides, user 'U18469' with a longer recommendation path $p_{\textit{longer}.u}$ is also involved in this experiment for a longer recommendation path case.  In this experiment, we utilize the CF algorithm to explore user belief degrees of categories. Figure \ref{fig:entropy_validate} describe the evolving levels of beliefs over time for the information that users initially perceive as extremely uninteresting based on their own belief networks.

\begin{figure}[htbp]
\centering
\subfigure[Temporal Variation in User Beliefs about Categories of Interest and Disinterest on MIND]{\includegraphics[width=0.49\textwidth]{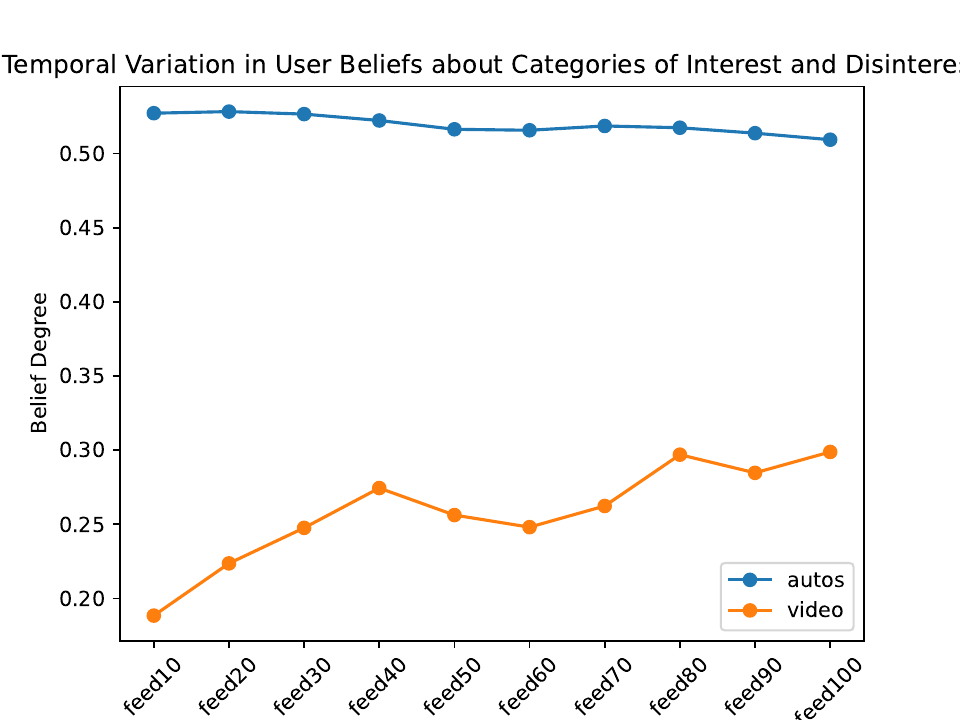}\label{fig:entropy_MIND}}
\subfigure[Temporal Variation in User Beliefs about Categories of Interest and Disinterest on IMDB]{\includegraphics[width=0.49\textwidth]{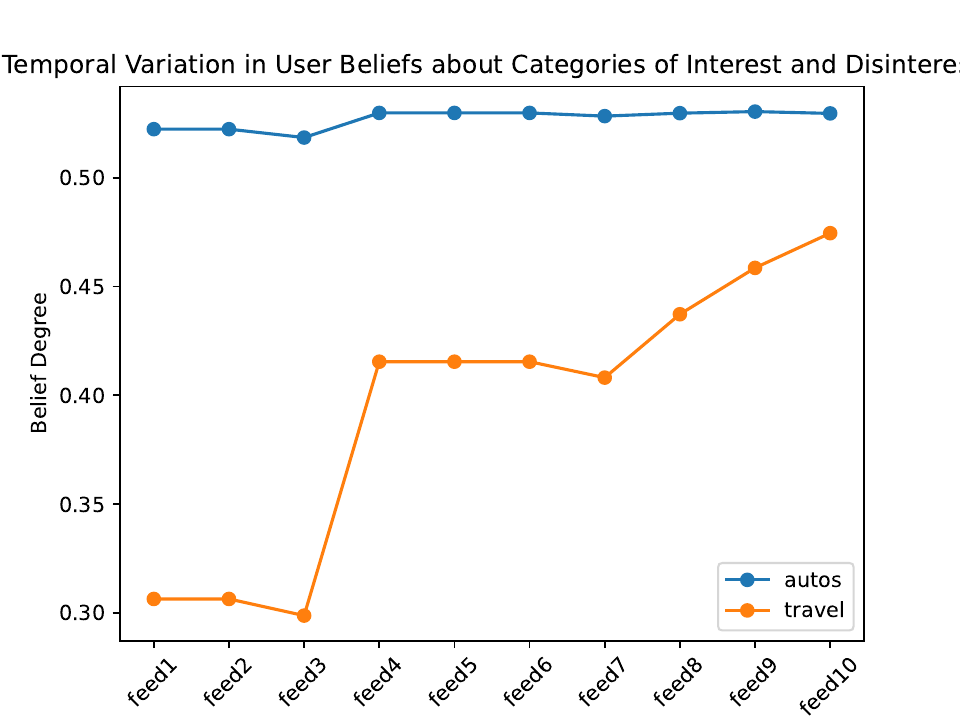}\label{fig:entropy_IMDB}}
\caption{Temporal Variation in User Beliefs about Categories of Interest and Disinterest}
\label{fig:entropy_validate}
\end{figure}

The BHEISR model is designed to leverage existing recommendation systems to present information to users, with the objective of achieving a balance within the user's belief network. Given that shifts in users' belief states require long-term guidance, we seek to demonstrate that users tend to receive less preferred information via recommendations in a relatively short timeframe. In pursuit of this objective, we selected user 'U276' from the IMDB dataset, which presents a shorter recommendation path $p_{\textit{shorter}.u}$ (indicating closer feature distances between categories of interest and disinterest). This strategy allows us to observe changes in user beliefs within a short period. Furthermore, user 'U18469', exhibiting a longer recommendation path $p_{\textit{longer}.u}$, is also included in this experiment to represent cases involving longer recommendation paths. This experiment employs the Collaborative Filtering (CF) algorithm to examine user belief degrees across various categories. Figure \ref{fig:entropy_validate} shows the temporal evolution of belief levels for information initially deemed uninteresting by the users, according to their belief networks.

%For the $"p_{\textit{longer}}"$ users, to present the trend of belief degree more clearly, in Figure \ref{fig:entropy_MIND}, we have magnified the time span to represent 10 feeds as the one-time interval. This means that we observe the users' level of interest in the "autos" and "video" categories only once every 10 recommendations. On the other hand, for the $"p_{\textit{shorter}}"$ users, we capture the cognitive changes for each recommendation. In the figure, "autos" represents a category that both users are interested in, "video" represents a category that the MIND user is not interested in, and "travel" represents a category that the IMDB user is not interested in.

For users with longer paths, $"p_{\textit{longer}}"$, we have adjusted the time scale in Figures \ref{fig:entropy_MIND} and \ref{fig:entropy_IMDB} to better illustrate the trend in belief degree: a single time interval now represents 10 recommendation feeds. Consequently, we measure the users' level of interest in the "autos" and "video" categories once every 10 recommendations. In contrast, for users with shorter paths, $"p_{\textit{shorter}}"$, we capture belief changes following each recommendation. In these two figures, "autos" represents a category of shared interest for both users, "video" refers to a category that the MIND user is disinterested in, and "travel" indicates a category of disinterest for the IMDB user.

%From both figures, we can observe that as time progresses, the users' affinity for "autos" remains stable or shows a slight downward trend. On the other hand, for the categories that the users initially showed no interest in, continuous recommendations have sparked their interest. Especially for the IMDB user, there is a significant increase in interest in the "travel" category. This strongly supports the effectiveness of the BHEISR-based model in achieving user cognitive harmony.

From both figures, we observe that the users' preference for "autos" either remains stable or displays a minor declining trend as time progresses. Conversely, for the categories that the users initially showed no interest in, continuous recommendations have sparked their interest. Particularly for the IMDB user, a significant increase in interest towards the "travel" category is evident. This observation robustly validates the efficacy of the BHEISR-based model in fostering belief harmony among users.

\subsubsection{Experiment 3: \textbf{Filter Bubble Users Detection}}

%In this experiment, we take all MIND and IMDB users affected by filter bubbles as the experimental subjects. This experiment is used to detect the change in the number of filter bubble-affected users over time under different recommendation models based on the CF algorithm.

In this experiment, we focus on all MIND and IMDB users influenced by the filter bubble effect, serving as our experimental subjects. The purpose of this experiment is to track the temporal variations in the number of filter bubble-impacted users under different recommendation models, all of which are based on the CF algorithm.

\begin{table}[htbp]
\centering
\renewcommand{\arraystretch}{1} % Increase row height
\resizebox{\textwidth}{!}{
\begin{tabular}{@{}llllllll|lllllll@{}}
\hline
\multicolumn{8}{c|}{\textbf{MIND}} & \multicolumn{7}{c}{\textbf{IMDB}} \\ \hline
\textbf{Times} & \textbf{RD} & \textbf{$RD_{w}C$} & \textbf{CB} & \textbf{$CB_{w}C$} & \textbf{UC} & \textbf{$UC_{w}C$} & \textbf{BHEISR} & \textbf{RD} & \textbf{$RD_{w}C$} & \textbf{CB} & \textbf{$CB_{w}C$} & \textbf{UC} & \textbf{$UC_{w}C$} & \textbf{BHEISR} \\ \hline
$feed_{1}$ & 28 & 28 & 2 & 2 & 28 & 24 $\downarrow$& 26 & 6 & 4 $\downarrow$ & 0 & 0 & 6 & 4 $\downarrow$ & 4\\\hline
$feed_{2}$ & 30 & 24 $\downarrow$ & 2 & 2 & 28 & 24 $\downarrow$& 26 & 6 & 4 $\downarrow$ & 0 & 0 & 6 & 4 $\downarrow$ & 4\\\hline
$feed_{3}$ & 30 & 26 $\downarrow$ & 2 & 2 & 28 & 24 $\downarrow$ & 26 & 6 & 4 $\downarrow$ & 1 & 0 $\downarrow$ & 6 & 4 $\downarrow$ & 4\\\hline
$feed_{4}$ & 30 & 26 $\downarrow$ & 4 & 4 & 28 & 24 $\downarrow$& 26 & 6 & 4 $\downarrow$ & 2 & 0 $\downarrow$ & 6 & 6 & 4\\\hline
$feed_{5}$ & 30 & 24 $\downarrow$ & 4 & 4 & 28 & 24 $\downarrow$& 26 & 6 & 5 $\downarrow$ & 3 & 1 $\downarrow$ & 6 & 6 & 4\\\hline
$feed_{6}$ & 30 & 24 $\downarrow$ & 4 & 4& 28 & 24 $\downarrow$& 26 & 6 & 5 $\downarrow$ & 3 & 2 $\downarrow$ & 6 & 6 & 4\\\hline
$feed_{7}$ & 30 & 24 $\downarrow$ & 4 & 4 & 28 & 24 $\downarrow$ & 26 & 6 & 5 $\downarrow$ & 3 & 3 & 6 & 6 & 4\\\hline
$feed_{8}$ & 30 & 24 $\downarrow$ & 6 & 4 $\downarrow$ & 28 & 24 $\downarrow$& 24 & 6 & 5 $\downarrow$ & 3 & 2 $\downarrow$ & 6 & 6 & 4\\\hline
$feed_{9}$ & 30 & 26 $\downarrow$ & 6 & 4 $\downarrow$ & 28 & 24 $\downarrow$& 24 & 6 & 5 $\downarrow$ & 3 & 2 $\downarrow$ & 6 & 6 & 4\\\hline
$feed_{10}$ & 30 & 26 $\downarrow$ & 10 & 4 $\downarrow$ & 28 & 24 $\downarrow$& 24 & 6 & 5 $\downarrow$ & 4 & 2 $\downarrow$ & 6 & 5 $\downarrow$ & 4\\\hline

\end{tabular}
}
\caption{Filter Bubble Users Detection on MIND and IMDB datasets.}
\label{tab:MIND_IMDB}%
\end{table}

%Table \ref{tab:MIND_IMDB} describes the number of filter bubble-affected users under different recommendation systems. From a horizontal perspective, the models based on BHEISR for recommendations perform better than the original models, with fewer users affected by the filter bubble. This demonstrates the effectiveness of BHEISR in mitigating the filter bubble and balancing user beliefs. Besides, except that the CB-related model has not changed much, the number of other users affected by the filter bubble has decreased significantly. From the table, we can also observe that comparing the BHEISR model as a recommendation model with the BHEISR model as an intermediate agency, the BHEISR model as an intermediate agency has better performance in alleviating filter bubbles. Especially on the MIND dataset with a relatively large amount of data.

Table \ref{tab:MIND_IMDB} describes the number of users affected by the filter bubble under various recommendation systems. From a horizontal perspective, recommendation models based on the BHEISR strategy outperform the original models, affecting fewer users with the filter bubble. This underlines the effectiveness of BHEISR in mitigating the filter bubble effect and establishing balance in user beliefs. Moreover, except for the minor changes observed in the CB-related model, there is a notable decrease in the number of users affected by the filter bubble across other models. The table further reveals that, when comparing the performance of the BHEISR model as a recommendation model to its use as an intermediate agency, the latter demonstrates superior capability in alleviating filter bubbles, particularly in datasets with substantial data volume such as the MIND dataset.

%From a vertical perspective, the number of filter bubble-affected users varies over time. For example, in the IMDB dataset, the $UC_{w}C$ model and the  $RD_{w}C$ model in the MIND dataset. They all show a fluctuation in the number of filter bubble users. The number of filter bubble users from IMDB, starting from 4, increasing to 6, and then decreasing to 5. Meanwhile,  The number of filter bubble users from MIND also shows fluctuant changes from 28 to 24, from 24 to 26, from 26 to 24, and finally from 24 to 26. This strongly indicates that users' awareness is influenced by the nudging technique and rescheduling path approach incorporated in the BHEISR model. The BHEISR model continuously adjusts the recommendation path based on user feedback to influence user behavior.

From a vertical perspective, the number of users affected by the filter bubble fluctuates over time. For instance, in the IMDB dataset under the $UC_{w}C$ model, and in the MIND dataset under the $RD_{w}C$ model, there are evident fluctuations in the count of filter bubble-impacted users. In the IMDB dataset, the number of such users initially starts at 4, increases to 6, and then reduces to 5. Similarly, in the MIND dataset, the count of filter bubble users exhibits variable changes, from 28 to 24, then from 24 to 26, next from 26 to 24, and finally from 24 to 26. These observations strongly suggest that user belief is influenced by the nudging technique and the rescheduling path approach integrated within the BHEISR model. The BHEISR model persistently adjusts the recommendation path based on user feedback, thereby influencing user behavior.

\subsubsection{Experiment 4: \textbf{Parameter Analysis}}

%In our experiments, we investigate the influence of different weight $w$ of BHEISR-generated items $GI$ within a recommendation feed on user belief diversity. We specifically analyze how varying the proportion of BHEISR-generated items affected the diversity of perspectives among users. By manipulating the $w$ parameter, we can examine the impact of different levels of intervention by the BHEISR model on recommendation outcomes. In this experiment, we use the user 'U1629' from the MIND dataset as an example. Besides, we select $UC_{w}BHEISR$ as the experimental model in this experiment.

This experiment aims to investigate the impact of varying weights $w$ of BHEISR-generated items $GI$ within a recommendation feed on user belief diversity. The specific focus is on analyzing how the diversity of perspectives among users is influenced by changing the weight of BHEISR-generated items. By manipulating the $w$ parameter, we can explore the effect of different degrees of intervention from the BHEISR model on the recommendation results. In this experiment, we utilize user 'U1629' from the MIND dataset as a representative subject. Additionally, we select $UC_{w}BHEISR$ as the experimental model for this experiment.

\begin{figure}[htbp]
		\centering
		\includegraphics[width=0.60\textwidth]{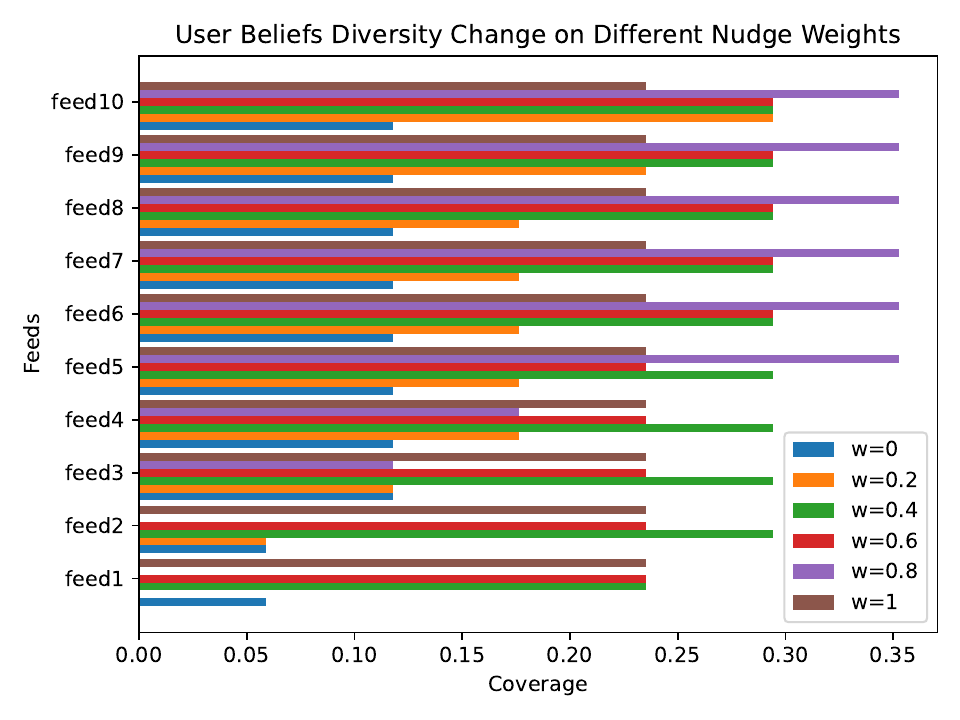}
    \caption{User Beliefs Diversity Change on Different Nudge Weights}
	 \label{fig:parameter_Users}
\end{figure}

%From this bar graph \ref{fig:parameter_Users}, it can be observed that the BHEISR recommendation weight has a significant impact on user diversity. It is evident that as the number of recommendations increases over time, the influence of the recommendation weight becomes more pronounced. Specifically, when the weight is 0.4, users have a higher probability of accepting the recommendations, and this influence is stable. However, after a certain number of recommendations, the impact of BHEISR-recommended information on users reaches its peak. The greater the proportion of items recommended by BHEISR in the total recommendation, the greater the impact on users in the later stage of recommendation.

As can be seen from Figure \ref{fig:parameter_Users}, the weight assigned to BHEISR recommendations significantly influences user belief diversity. It becomes clear that as the number of recommendations increases over time, the impact of the recommendation weight grows more pronounced. Notably, when the weight is set at 0.4, users show a higher probability of accepting the recommendations, and this influence remains steady. However, after a certain number of recommendations, the impact of BHEISR-recommended information on users reaches its peak. The larger the fraction of items recommended by BHEISR in the total recommendation list, the greater pronounced the impact on users in the later stages of recommendations.

In conclusion, the analysis indicates that the weight of BHEISR recommendations considerably affects user belief diversity. The influence of this weight amplifies as the number of recommendations increases. However, there is a saturation point at which the impact of BHEISR-recommended information on users reaches its maximum. In the experiment, we select $w$=0.6 as the BHEISR recommendation weight owing to its consistently improving performance.

\section{Conclusion and Future Work}

In this research work, we focus on addressing the issue of filter bubbles in recommendation systems and propose the BHEISR model as a solution, with the objective of mitigating the negative effects of filter bubbles and promoting belief harmony among users.

The BHEISR model functions as an intermediary between existing recommendation systems and users, aiming to facilitate democratic and transparent recommendations. It incorporates several key features. First, it utilizes the Filter Bubbles Detection Model based on Multi-faceted Reasoning for filter bubbles identification. Second, it applies nudging techniques to incrementally broaden users' interests and balance their beliefs. Finally, it incorporates a user feedback loop based on Generative GAI to capture evolving user beliefs over time and increase recommendation diversity.

The experimental results explicitly reveal the effectiveness of the BHEISR model in mitigating filter bubbles and balancing user beliefs. Real-world datasets and nearly 200 filter bubble-affected users were used to validate the model's performance. 

As for future research work, we plan to explore additional techniques to augment the model's performance further, undertake user studies to assess the long-term impacts, and investigate the influence of the BHEISR model on critical thinking abilities and inclusivity. Persistent refinement and evaluation of the model could lead to improved recommendation systems that cater more effectively to the diverse needs of users.

\section{Acknowledgments}

The authors would like to acknowledge the financial support from Callaghan Innovation (CSITR1901, 2021), New Zealand, without which this research would not have been possible. We are grateful for their contributions to the advancement of science and technology in New Zealand. The authors would also like to thank CAITO.ai for their invaluable partnership and their contributions to the project.

\bibliographystyle{ACM-Reference-Format}
\bibliography{bibFile}
\end{document}